\documentclass[journal,12pt,onecolumn,draftclsnofoot]{IEEEtran}
\usepackage{times}
\usepackage{amsmath,epsfig,amssymb,mathrsfs}
\usepackage{graphics}
\usepackage{array}
\usepackage{subfig,keyval,ragged2e,everysel}
\usepackage{graphicx}
\usepackage{caption}
\usepackage{amsthm}
\usepackage{float}
\usepackage{amssymb}
\usepackage{moreverb,algorithm,algorithmic}
\usepackage{fancybox}
\usepackage{cite}
\usepackage{balance}
\usepackage{lipsum}
\usepackage{colortbl}

\newcommand{\rp}[1]{\iftrue{\color{cyan}~[RP: #1]}\fi}

\begin{document}
\title{In-network Collaboration for CDMA-based Reliable Underwater Acoustic Communications}
\author{{Mehdi Rahmati,~\IEEEmembership{Student Member,~IEEE,} Roberto Petroccia,~\IEEEmembership{Senior Member,~IEEE}, and Dario Pompili,~\IEEEmembership{Senior Member,~IEEE}}
\thanks{\IEEEcompsocthanksitem M.~Rahmati and D.~Pompili are with the Dept. of Electrical and Computer Engineering, Rutgers University--New Brunswick, NJ, USA. R.~Petroccia is with the NATO STO Centre for Maritime Research and Experimentation~(CMRE), La Spezia, Italy. \protect\\ 
E-mails: mehdi\_rahmati@cac.rutgers.edu, pompili@rutgers.edu; roberto.petroccia@cmre.nato.int\protect
\IEEEcompsocthanksitem A shorter version of this work is in the \emph{Proc. of IEEE Underwater Communications Networks~(UComms)}, Sept'18~\cite{rahmati2018collaborative}.\protect
\IEEEcompsocthanksitem This work was supported by the NSF CAREER Award No.~OCI-1054234 and by the NATO Allied Command Transformation~(ACT) Future Solutions Branch under the Autonomous Security Network Programme.  
 
 © 20XX IEEE.  Personal use of this material is permitted.  Permission from IEEE must be obtained for all other uses, in any current or future media, including reprinting/republishing this material for advertising or promotional purposes, creating new collective works, for resale or redistribution to servers or lists, or reuse of any copyrighted component of this work in other works.
}}

\clearpage\maketitle
\thispagestyle{empty}

\maketitle


\begin{abstract}
Achieving high throughput and reliability in underwater acoustic networks for transmitting distributed and large volume of data is a challenging task due to the bandwidth-limited and unpredictable nature of the acoustic channel. In a multi-node network, such as in the Internet of Underwater Things~(IoUT), communication link efficiency varies dynamically: if the channel is not in good condition, e.g., when in deep fade, channel coding techniques may fail to deliver the information even with multiple retransmissions. Hence, an efficient and agile collaborative strategy is required to allocate appropriate resources to the communication links based on their status.
The proposed solution adjusts the physical- and link-layer parameters collaboratively for a Code Division Multiple Access~(CDMA)-based underwater network. An adaptive Hybrid Automatic Repeat Request~(HARQ) solution is employed to guarantee reliable communications against errors in poor links. Results were validated using data collected from the LOON testbed---hosted at the NATO STO Centre for Maritime Research and Experimentation~(CMRE) in La Spezia, Italy---and from the REP18-Atlantic sea trial conducted in Sept'18 in Portuguese water.
\end{abstract}

%
\IEEEpeerreviewmaketitle

\section{Introduction}\label{sec:intro}

\textbf{Overview:}
Over the past decade, Underwater Acoustic Networks~(UANs) have attracted the attention of researchers, engineers, and practitioners, as they enable a wide range of applications such as oceanographic data collection, offshore exploration, tactical surveillance, pollution and noise monitoring, disaster prevention, and assisted navigation~\cite{rahmati2017unisec}. These networks face various challenges due to the unique and harsh characteristics of the propagation of underwater acoustic waves~\cite{pompili2009overview,Stojanovic2009,HeidemannSZ12}. In applications as the Internet of Underwater Things~(IoUTs), data is usually distributed across a high number of nodes, while a single node (sink) is used for data collection, fusion, and processing~\cite{petrioli2014sunrise}. The efficiency of an IoUTs system in mission-critical applications relies on the robustness of the communication algorithms and protocols that control the components of such a system. To maximize the achievable throughput of such a network, one of the major challenges is the design of a secure, robust, and scalable Medium Access Control~(MAC) and an Error Control~(EC) strategy. These solutions need in fact to guarantee 
low channel access delay, low energy consumption, and fairness among competing and/or collaborating nodes in the face of the harsh characteristics of the underwater acoustic propagation medium~\cite{PetrioliPS08,shahabudeen2014analysis,rahmati2015interference,PetrocciaPP18}.

\textbf{Motivation:}
Terrestrial and conventional MAC/link-layer communication techniques fail to provide the required robustness and reliability for futuristic applications due to the characteristics of the underwater acoustic channel~\cite{pompili2009overview,rahmati2018PSDMA}. Direct-Sequence Spread Spectrum Code Division Multiple Access~(DSSS-CDMA) is a promising physical-layer and multiple-access techniques for UANs since i)~it is inherently robust to frequency-selective fading, ii)~it compensates for the effect of multipath at the receiver by using filters that can collect the transmitted energy spread over multiple paths, and iii)~it allows receivers to distinguish among signals simultaneously transmitted in the same frequency band by multiple devices~\cite{pompili2009cdma,stojanovic2006multichannel,tsimenidis2001underwater}. The use of an efficient CDMA scheme, supporting an adaptive EC strategy such as Hybrid Automatic Repeat Request~(HARQ), has the potential to increase channel reuse and to reduce the number of packet retransmissions, thus increasing network reliability and achievable throughput, while decreasing the network energy consumption. However, since the number of retransmissions is limited in the practical truncated ARQ/HARQ EC strategies, the receiver might start dropping packets, thus significantly limiting the capability of delivering data in the network.

\textbf{Contribution:}
In this work, we extend the concept of point-to-point HARQ to an implicitly collaborative scenario in combination with a DSSS-CDMA approach. A transmitting node with low-quality communication links piggybacks on its neighboring nodes' transmissions when protecting its data against errors, in order to increase the system throughput. We propose a solution to achieve the following objectives: i)~high network reliability and throughput by allocating an appropriate share of system resources to different nodes; ii)~latency problem alleviation caused by the conventional HARQ retransmission strategy; iii)~simultaneous transmission on the available bandwidth via easily- and locally-generated CDMA chaotic codes using a secret seed with a flexible and large family size; and iv)~low energy consumption via efficient output power allocation. Data collected using the CMRE LOON testbed~\cite{alves2014loon} were investigated to validate the proposed method. This testbed is hosted in the Gulf of La Spezia, Italy, close to the CMRE premises and it is characterized by \emph{shallow-water} communications (occurring at a maximum depth of $15~\rm{m}$), which may be heavily affected by multipath. Our solution was able to find dynamically the optimal trade-off among these four objectives according to the application requirements. To evaluate the protocol performance under different conditions, additional data was then collected in a \emph{deep-water} scenario during the REP18-Atlantic sea-trial. This trial was organized by CMRE, the Portuguese Navy~(PRT-N), and the Faculty of Engineering of the University of Porto~(FEUP) in Portuguese water, between Sines and Sesimbra, in Sept'18.

\textbf{Article Organization:}
The remainder of this article is organized as follows. In Sect.~\ref{sec:rel}, we present
a summary of prior work on different acoustic data transmission techniques in the underwater environment, and CDMA is discussed as a candidate for the UANs. In Sect.~\ref{sec:Probsol}, we define the required parameters for the proposed solution, and then present the proposed collaborative hybrid ARQ solution to achieve reliable communications in underwater CDMA networks. Then, in Sect.~\ref{sec:Eval}, we provide experimental and simulation results along with observations based on data collected via the CMRE LOON testbed and during the REP18-Atlantic sea trial. Finally, in Sect.~\ref{sec:Con}, we conclude the article and provide insights into our future plan.

\section{Related Work}\label{sec:rel}

Conventional ARQ, as a feedback-assistant EC technique, requests a retransmission for the erroneously received data packet. When the error is detectable, the packet is discarded until the same packet is successfully received in the next round.
Retransmission is an appropriate solution to achieve a certain level of reliability in the underwater channel, specially when the Forward Error Correction~(FEC) schemes are not able to correct the burst errors alone. On the other hand, because of the long propagation delay in underwater channels, the performance drops significantly since a technique such as stop\&wait and other similar ARQ techniques fail to provide a reasonable throughput. Furthermore, having a feedback link might not be feasible in some practical systems or might be erroneous if it is available. Therefore, to reduce the number of retransmissions and to increase the system reliability under poor channel conditions, a powerful FEC code should be used, which makes the decoding hard to implement~\cite[ch.22]{costello2004error}.

In practical experiments---when usually the channel is error prone and therefore unreliable---multiple rounds of retransmissions should be performed to deliver the intended data; consequently, a huge amount of time is wasted given the long propagation delay in the underwater channel. Therefore, a proper combination of the ARQ and FEC is required in an efficient scheme to overcome the mentioned problems.
This combination of ARQ and FEC leads to a Hybrid approach, i.e., HARQ, which reduces the number of packet retransmissions and increases the system reliability, specially under poor channel conditions. If the data is not decodable, the receiver sends back a Negative Acknowledgement~(NACK) to the transmitter and asks for additional/duplicated FEC, which eventually increases the probability of successful transmission~\cite{soljanin2004hybrid}. However, if the channel is very noisy, even using multiple retransmissions may not work. In the truncated ARQ/HARQ, the number of retransmissions is limited. Therefore, the receiver might drop the data, which detrimentally affects the throughput of the network. 

A type-I~HARQ discards the erroneous received packet after a failed attempt to correct it, then the transmitter repeats the same packet until the error is corrected. This method might be inefficient in time-varying underwater acoustic channel. When the channel is in good condition, i.e., retransmission is not required, FEC information is more than it requires and so the throughput drops. On the other hand, if the channel is not in good condition, e.g., when in deep fade, the pre-defined FEC might not be adequate and the throughput drops again because of multiple retransmissions~\cite{stojanovic2005optimization}. 

A type-II~HARQ requires a larger buffer size and has a higher complexity and efficiency compared to type-I. It adapts itself with the channel in such a way that it first transmits the packet along with the error detection bits---similar to one of the ARQ schemes---when the channel is good. While the channel becomes worse and after detecting the erroneous packet, a NACK message is sent back and---rather than retransmitting the same packet as type-I does---FEC is transmitted to help decode the stored packet in the receiver's buffer. If the error persists, the second NACK is issued and the same FEC might be retransmitted or extra FEC might be added depending on the coding strategy. Incremental Redundancy~(IR)~HARQ, which shows a higher throughput efficiency in terrestrial time-varying channels, adds extra redundant information in each round of retransmission after receiving the NACK message~\cite{soljanin2004hybrid}. Terrestrial standards such as in High Speed Packet Access~(HSPA) and Long Term Evolution~(LTE) have exploited HARQ synchronously for the uplink, and asynchronously in the downlink direction. Authors in~\cite{pedersen2017rethink} discuss the requirements for designing a user-centric and network-optimized HARQ for the fifth generation~(5G) of mobile
communications. Given the necessity of supporting futuristic applications such as in IoUT, we believe that a new design for HARQ is essential.   

Using numerical simulations, authors in~\cite{ahmed2018grouped} used the random linear packet coding to control the 
packet loss in a hierarchical definition of packets in the stop\&wait ARQ protocol for the channels with a long propagation delay. In~\cite{casari2008towards}, the authors applied fountain codes to HARQ in underwater networks to reduce retransmissions and achieve optimal broadcasting policies. An adaptive coding approach based on the IR-HARQ was proposed in~\cite{diamant2015adaptive} to improve the packet error rate in a time-slotted underwater acoustic network. In~\cite{rahmati2015uwmimo}, we proposed a scheme based on HARQ that exploits the diversity gain offered by independent links of an underwater acoustic Multiple Input Multiple Output~(MIMO) channel. A large number of papers can be found in the literature that investigate the efficiency of point-to-point HARQ, especially in the terrestrial environment.

Various works have been proposed addressing separately CDMA and HARQ for UANs. Authors in~\cite{stojanovic2006multichannel} discuss DSSS CDMA as a candidate MAC for mobile UANs in which multiple nodes connect to a central receiver. A distributed single-carrier CDMA underwater MAC was proposed in~\cite{pompili2009cdma}, which aims at achieving high network throughput, low channel access delay, and low energy consumption. Although Pseudo-Noise~(PN) codes have been extensively employed in spread spectrum communication systems, considering their limitation in the number of different PN sequences and their cross-correlation properties, \cite{heidari1994chaotic} proposed using chaotic sequences in DSSS communications. These sequences can be generated through an uncomplicated deterministic map. Moreover, since chaotic systems are extremely dependent on the initial conditions, they can produce an infinite set of orthogonal uncorrelated sequences. One widely studied chaotic set that has been employed for underwater communications~\cite{azou2002chaotic} is generated based on the Logistic map. This map is able to produce a variety of distinct sequences for different users, by just changing the initial states and/or its bifurcation parameter. The proposed algorithm in~\cite{pompili2009cdma} uses locally-generated chaotic codes to spread transmitted signals on the available bandwidth, which guarantees secure protection against eavesdropping (as packets can only be decoded with the proper chaotic code, which depends on the secret initial conditions/seed), transmitter-receiver self-synchronization, and good auto- and cross-correlation properties~\cite{Broomhead99}.

\section{Problem Definition and Proposed Solution} \label{sec:Probsol}

\textbf{Problem Definition:}
While HARQ is a reliable EC solution based on packet retransmission, in practical scenarios and in the truncated HARQ, the number of retransmissions is limited. Therefore, if the underwater channel quality cannot be guaranteed, then multiple retransmissions may be not sufficient to correctly deliver the intended data, thus detrimentally affecting the reliability and the total throughput of the network. The problem gets worse when the interference from other users is involved in the performance. Therefore, a solution should be provided especially for multiuser UANs. CDMA is a promising technique for UANs since it can provide the required robustness and security of low-data-rate communications in multiuser scenarios in which all the nodes can overlap at the same time and at the same frequency band---despite the limited bandwidth---without any interference. Although M-sequences are popular in many CDMA systems, their cross-correlation shows some partial correlation for larger length of sequences in the multipath channels~\cite{heidari1994chaotic}. Moreover, the number of users that can be supported is limited by a sequence, which is a serious restriction for IoUT scenarios.

\begin{figure}[!t]
\centering 
\includegraphics[width=8.4cm
]{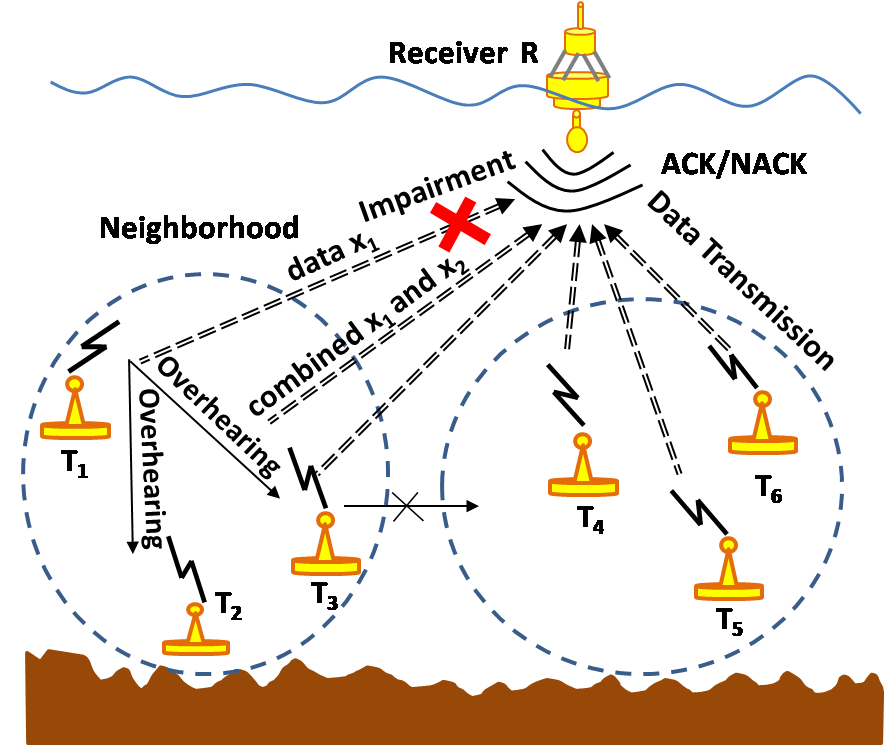}
\caption{ \small Architecture showing transmitting nodes,
$T_i, i=1,...,N_m$ in neighborhood $m,\; m=1,...,M$, when the channel quality varies from one link to another. As an example, data from node $T_1 \in N_1$ fails to reach the receiver. CDMA is exploited as the multiple access technique. The nodes overhear and collaborate in the HARQ procedure based on their communication links quality to improve the system throughput.}
\label{fig:systemmodel}
\end{figure}

\textbf{Proposed Architecture:}
Figure~\ref{fig:systemmodel} shows the case in which $N$ transmitting nodes, $T_i,\;i=1,2,...,N_m$, in neighborhood areas $m=1,...,M$ combine independently-sensed information, $x_i(r)$ (from independent nodes) for data fusion at the receiver $R$ (the sink) at different transmission rounds $r=1,2,...,r_T$. Nodes in the same neighborhood can overhear other transmissions and collaborate to deliver the intended data, similarly to what $T_2$ is doing for $T_1$ in Fig.~\ref{fig:systemmodel}. 

We use chaotic sequences in CDMA to increase the security in the communications. Albeit deterministic, chaotic codes look like noise, similarly to PN sequences; however, they are different for every bit of transmitted data. Hence, it is much harder for an eavesdropper, i.e., an unauthorized node outside of the neighborhood without the knowledge of the used codes (seed plus generating map), to regenerate the sequences and extract the data. This property allows us to have authorized nodes collaborating in a more secure way in the defined scenario and to also guarantee the service to a large number of users. Nodes that need to exchange data have to share the same family/seed of the code, with the chaotic codes generated in a deterministic way once the family/seed is known. This concept of sharing the same family/seed of the code is similar to the sharing of an encryption key in the context of symmetric cryptography. Before deploying the network, nodes are assigned a key or set of keys to talk to the other nodes. Nodes sharing the same key can read each others data. The use of more keys enables to create different cluster of nodes capable of sharing data. Different keys can be also used for different nodes or type of messages, e.g., some message may have higher priority or classification level and need to be shared with (or handled by) a reduced number of nodes so a specific key has to be used. Not only chaotic sequences provide the security in the channel, but they also show a considerable robustness against the multipath effect due to their good auto-and cross-correlation functions. Yet, CDMA requires to optimize the transmit power and spreading code length to limit the near-far problem and to maximize system throughout.

\textbf{CDMA-based Collaborative HARQ:}
The proposed collaborative HARQ for data protection, combined with a CDMA (using chaotic codes) for secure and interference-free transmissions, relies on a closed-loop strategy based on measurements sent back by the receivers. This is to avoid relying on the unrealistic symmetric-link assumption, which does not usually hold in the underwater environment. Each receiver periodically collects information on the channel state. This information is then provided to the neighbors by transmitting short ACK/NACK messages.

For each neighborhood, the error probability on the decoded sequence $\tilde{x}_i$ for the transmitted codeword ${x}_i$ from node $T_i$ can be upper-bounded using the \emph{Bhattacharyya bound}~\cite{john2008digital,soljanin2004hybrid} as, $P_{e}\big(x_i,\tilde{x}_i\big)\leqslant {B}_i^{h}$,
where $h$ is the Hamming distance and $B_i$ is the Bhattacharyya Parameter~(BP) in a noisy channel. Here we assume that channel does not change during one transmission round. This parameter is defined for every transmitted bit $x$ and received bit $y$ as $B=\sum_{y\in\Omega} \sqrt{\Pr(y|x=0)\Pr(y|x=1)}$. Here, $\Omega$ stands for the output alphabet and $\Pr(y|x=0)$ and $\Pr(y|x=1)$ are transition probabilities, $\forall y \in \Omega$.
This parameter is considered as a channel reliability metric as it is an upper-bound on the probability of error in a typical Maximum-Likelihood~(ML) detection problem, where larger BP values suggest channel unreliability and viceversa. The union-Bhattacharyya bound~\cite{soljanin2004hybrid} can be calculated for each channel for node $T_i$ as,
\begin{equation}\label{eq:union-bhatta}
{P^c_{e\:i}}\leqslant\sum_{h'=1}^n A_{h'} {{B_i}}^{h'},
\end{equation}
where ${P^c_{e\:i}}$ denotes the codeword error probability of code $c$ from family code $\mathbb{C}$ and $A_{h'}$ represents the codewords with weight $h'$. Data is encoded using a pre-defined mother code. Data and the first portion of parity bits are transmitted in the first round. If the receiver cannot decode the data, then a NACK will trigger the transmitter to send the second portion of parity bits in the next round, so that the receiver possibly can decode with the help of both portions. We transmit the coded data in $(r_T)$ transmission rounds from the selected nodes based on Algo.~\ref{algo:AHARQalgo}, which will be discussed later.  

For each round of transmission ($r$), the error probability of the transmitted packet from node $T_i$, in a DSSS-CDMA system, is upper-bounded as~\cite{john2008digital}, 
\begin{equation} \label{eq:cdma1}
{P_{e}}_i (r)\leqslant (2^k-1) \cdot Q\Bigg(2\sqrt{\frac{P_{i}}{J_i}SL_{i}\;{R_{ci}}\;d_i}\Bigg),
\end{equation}
where $P_i$ is the transmitting power of $T_i$, $Q\big(.\big)$ is the Q-function, $J_i$ is the total interference and noise experienced by $T_i$, $R_{ci}=k/n$ is the coding rate of a code $c(n,k)$, ${d_i}$ is the minimum hamming distance~($h_i$), and $SL_{i}$ is the length of CDMA spreading code. Note that \eqref{eq:cdma1} implies that the transmitted power, amount of interference, rate, strength of channel coding, and the processing gain of CDMA system, all affect the probability of error. 

The maximum achievable spectral efficiency achieved by each node in a neighboring area $m$, at round $r$ is,
\begin{equation} \label{rate}
R_i(r)=\log \Bigg(1+\frac{\alpha_iP_ig_i}{N_0W+\sum_{j=1,j\neq i}^{N_m} \alpha_j P_j g_j+\mathbb{J}_m}\Bigg),
\end{equation}
where $W$ is the channel bandwidth, $N_0$ is the noise Power Spectral Density~(PSD), and $g_i$ is the channel gain. The transmit power is controlled by $P_i$ and $\alpha_i$ in each round of transmission. $\mathbb{J}_m$ is the interference from other neighborhoods of $m$ defined as,
\begin{equation}
    \mathbb{J}_m=\sum_{k=1,k\neq m}^{M}\sum_{j=1}^{N_k} \alpha_j P_j g_j.
\end{equation}
Note that $\sum_{j=1,j\neq i}^{N_m} \alpha_j P_j g_j+\mathbb{J}$ is the total interference that is undesirable from the CDMA perspective; however, we leverage the first term in our proposed HARQ to engage other nodes in the same neighborhood to collaborate in the process.

The other promising metric is the long-term throughput, which is defined based on the renewal reward theorem~\cite{zorzi1996use} as, $\eta={\mathbb{E}[\tilde{X}]}/{\mathbb{E}[\tilde{T}]}$. Hence, $\mathbb{E}[.]$ defines the expectation of a random variable, $\mathbb{E}[\tilde{X}]=X\big(1-\overline{\Pr_{out}}\big)$ is the number of decoded information nats, i.e, natural information unit. $\overline{\Pr_{out}}$ can be defined as the probability that the data has not been decoded after $(r)$ rounds. $\mathbb{E}[\tilde{T}]$ is the number of attempts for channel use during a packet transmission period.
We decrease $\overline{\Pr_{out}}$ via node collaboration and by adjusting the corresponding parameters so as to improve the network long-term throughput. The probability of decoding in round $(r)$ given that the data has not been decoded in the previous $(r-1)$ rounds is equivalent to $\Pr\big(\rm{NACK}_1,...,\rm{NACK}_{r-1},\rm{ACK}_r \big)$.
In our proposed method, when an impaired node gets its first NACK, we prevent getting more NACKs via the help of the collaborating nodes. Therefore, the average number of transmissions after ($r_T$) rounds for an impaired node $i$ can be calculated as,
\begin{equation}
\big(1-{P_{e}}_i(r=1)\big)+\sum_{r=2}^{r_T}\Big[ r(1-q ){P_{ei}^{r-1}}(r)\big(1-P_{ei}(r)\big)+rq{P_{e\kappa}^{r-1}}(r)\big(1-P_{e\kappa}(r)\big)\Big],
\end{equation}
where $q$ is the probability of collaboration and $\kappa$ is the collaborating node.

\textbf{Total Rate Maximization:}
In a multi-node system, in which nodes are influenced by each others' activities and are affected by the collaborators' coding scheme, those---which are involved in throughput and rate---should be selected in an optimal way. These parameters can be discussed under two major constraints, as follows.

\textit{Signal-to-Interference-Noise-Ratio~(SINR) Constraint:}
To find the constraint for the multiuser interference in a CDMA system, we should reassure that a minimum required SINR and so the minimum error rate is satisfied at the receiver. This parameter---as a popular metric for the Quality of Service~(QoS)---is a factor of processing gain, coding gain, and the signal power to interference ratio. 
Processing gain in CDMA represents the gain that is obtained by expanding the bandwidth of the signal and is shown by the spreading length~\cite{john2008digital}. Performance of the channel coding is decided by its coding gain and the minimum hamming distance.     

\textit{Power Constraint:}
When we amplify the transmit power, the received SNR will be improved; however, it causes more interference to the other nodes. We try to regulate the transmit power and to reduce the interference to the other neighborhoods by a power control strategy. The peak transmitting power of each node in every neighborhood should be bounded to a pre-defined maximum power $P_{max}$, i.e., $P_i\in [0,P_{max}]$. As a result of the interference, SINR, and channel impairment, a power-control coefficient $\alpha_i$ is decided in each round that matches the HARQ procedure.

\textit{General Optimization Problem:}
To maximize the total rate and to satisfy the performance and power constraints, we cast an optimization problem to find the optimum parameter vector $\Theta=[\theta_1,...,\theta_{N_m}]$, where $\theta_i=[P_i,R_{ci},d_i,SL_{i},\alpha_i]$ and $\;i=1,...,N_m$.
\begin{subequations}\label{opt}
\begin{align}
&\mathop {\max }\limits_{\Theta} \;  \mathcal{F}(r)=\sum_{i=1}^{N_m} R_{ci}R_i(r) \label{opt_1} \\ 
&\text{s.t.} \hspace{0.3cm} \rm{SINR\: constraint:} \gamma_i(r) =\big(2SL_i\big)_{dB}+\big(R_{ci}d_i\big)_{dB}+\big( \frac{\alpha_i P_{i}g_i }{J_i+\mathbb{J}_m} \big)_{dB}
\geq \gamma_{min},\label{opt_3}\\
& \hspace{0.7cm}
\rm{Power\: constraints:}
\sum_{i=1}^{N_m} \alpha_i P_i g_i \leq P_{th},\\
& \hspace{4.1cm}
 P_i- P_{max} \leq 0, \; \; \; i=1,2,...,N_m \\
 & \hspace{4.1cm}
 \sum_{i=1}^{N_m}  \alpha_i \leq N_m, \; \; \; \; \alpha_i \in \{0,1\},
\end{align}
\end{subequations}
where $\gamma_i(r)$, in $dB$, is the received Signal-to-Interference-Noise-Ratio~(SINR) from $T_i$ at round $r$, $J_i=N_0W+\sum_{j=1,j\neq i}^{N_m} \alpha_j P_j g_j$, and $\gamma_{min}$ is the minimum SINR, which is proportional to the probability of error in HARQ and determines the level of performance. $P_{th}$ guarantees that the total received power in $m$ does not affect other neighborhoods.   

Let $\Psi=[\psi_1,...,\psi_L]$ be the vector of Lagrange multipliers and $L=2N_m+2$ be the number of constraints. We form the Lagrangian function as $\mathcal{L}(\Theta, \Psi)= \mathcal{F}-\sum_{l=1}^{L} \psi_{l}(g_{l}(\Theta)-b_{l})$,
where each $g_{l}(\Theta)$ and $b_l$ are determined by each constraint such that $g_{l}(\Theta) \leq b_l$. 
\begin{equation}
\begin{split}
\mathcal{L}(\Theta , \Psi)=\mathcal{F}-\sum_{l=1}^{N_m} \psi_{l}\:(P_l- P_{max})-\sum_{l=1}^{N_m} \psi_{l+N_m}\:(-\gamma_l+\gamma_{min})\\
-\psi_{2N_m+1}\:(\sum_{l=i}^{N_m} \alpha_i P_i g_i - P_{th})-\psi_{2N_m+2}\:(\sum_{i=1}^{N_m}  \alpha_i-N_m).
\end{split}
\end{equation}

To find the optimum values, using Kuhn-Tucker condition, $\bigtriangledown_{\Theta} \mathcal{L}(\Theta , \Psi)=0$ should be solved.
%
There are $L$ complementary equations that should be held as $\psi_{l}\big(g_{l}(\Theta)-b_{l}\big)=0,\;l=1,...,L$, such that $\psi_l \geq 0 $. The feasible results of these equations determine the optimum parameter that results in maximum spectral efficiency. A numerical solution is presented in Sect.~\ref{sec:Eval} for this problem based on the experimental data collection. 

\begin{figure*}[!t]
\centering 
\includegraphics[width=16.5cm
]{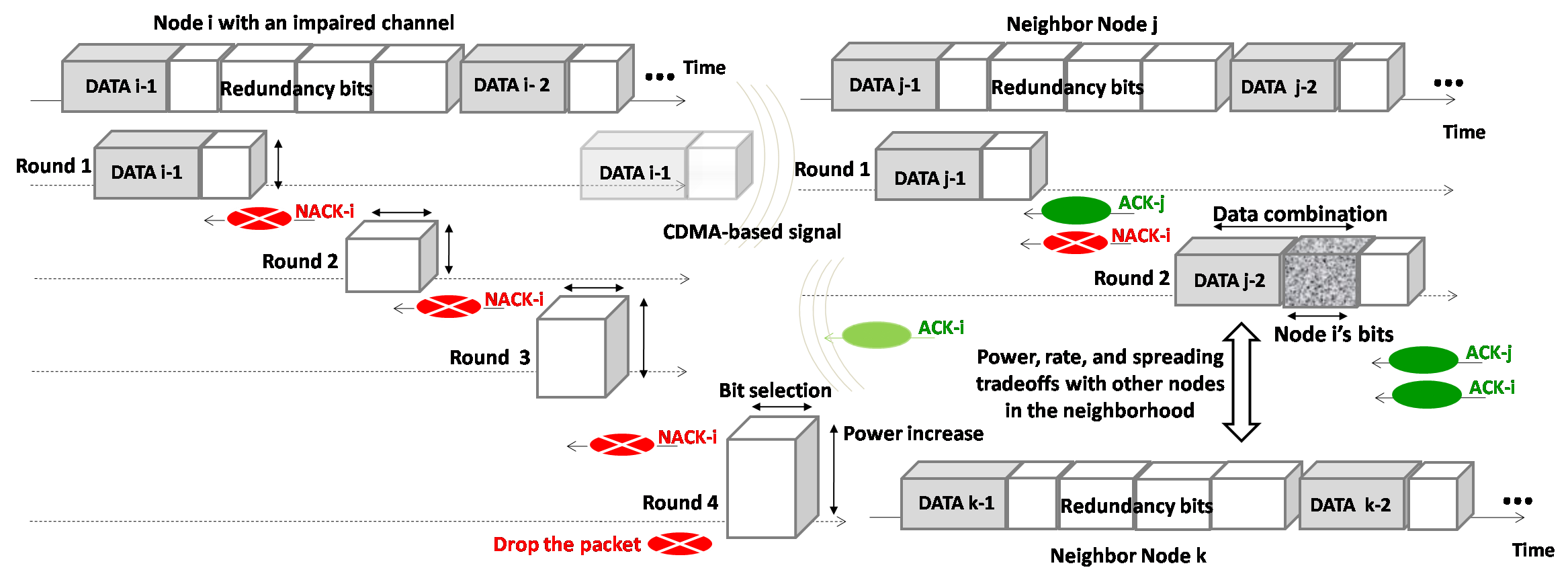}
\caption{\small Proposed protocol showing the interaction of the node with impaired channel and a neighbor. Without using this protocol~(left side of the figure), node $i$ would keep sending incremental redundancy and the packet would drop after $4$ rounds, while using our collaborative CDMA-based method (right side of the figure), $2$ rounds are sufficient for delivering the data; therefore, it reduces the end-to-end~(e2e) delay and increases the total throughput. The nodes decide how many bits to select, i.e., data rate, spreading parameters, and transmission power navigating various trade-offs collaboratively with the other active nodes in the neighborhood.}\label{fig:protocol_2}
\end{figure*}

\begin{algorithm}[!t]
\caption{Collaborative HARQ for nodes $T_i \in \mathcal{N}_m$ in the same neighborhood} \label{algo:AHARQalgo}
\small
\begin{algorithmic}[1] \small
\STATE {$\forall i =1,..., N_m:\;  r_i\leftarrow1$ as the index of transmission round; define $r_T$}
\STATE{choose collaborators $\hat{c}$ based on Eq.~\eqref{eq:union-bhatta} and the neighbor discovery algorithm; share the chaotic map with $\hat{c}$} 
\IF{event \big(request to send for $T_i \in \mathcal{N}_m$ \big)}
\WHILE{time-out codeword(\;)}
\STATE{$\forall T_i$: Make the packets considering Eq.~\eqref{eq:union-bhatta}; HARQ block formation}
\STATE {generate the chaotic code; solve problem in Eq.~\eqref{opt} considering Eq.~\eqref{eq:cdma1}}
\STATE{puncture the codeword; HARQ transmission procedure ( )}
\STATE{event (wait for $\rm{ACK_i/NACK_i}$)}
\WHILE{ $\rm{NACK_i}$ AND $r_i \leqslant r_T$}
\STATE {$r_i\leftarrow r_i+1$; construct the codeword with additional redundancy; rate and power selection}
\STATE{HARQ Retransmission procedure(\;)}
\ENDWHILE
\IF {$\rm{ACK_i}$}
\STATE{goto end}
\ELSE
\FOR {$ \forall\;T_{\widehat{i}}(r_T)$ (impaired nodes)}
\STATE{$\alpha_{\widehat{i}} \leftarrow 0$; revise the collaborators $j \in \hat{c}$ based on Eq.~\eqref{eq:union-bhatta} AND the received $\rm{NACK_{\widehat{i}}}$ AND $\rm{ACK_{j \neq \widehat{i}}}$ }
\FOR {$j=1:\hat{c}$}
\STATE{solve Eq.~\eqref{opt} for new $\Theta^*_j$ for all $\hat{c}$}
\IF{$\rm{arg\;max}\; \{R_j\}_{j\in \hat{c} \;\& \;j \neq \widehat{i}}>R_{\widehat{i}}$}
\STATE{\textbf{repeat} steps 3-12 for \textit{combined} data of $j$ and $\hat{i}$ until {$\rm{ACK_j}$} OR $r_T$\;\; \% \emph{the new collaboration}}
\ELSIF {$R_{\widehat{i}}< \sum R_{\widehat{c}}$}
\STATE{update $\sum R_{\widehat{c}}$ with the next collaborator $\rm{arg\;max}\; \{R_j\}_{j\in \hat{c} \;\& \;j \neq \hat{c}(max)}$}
\STATE{\textbf{repeat} steps 3-12 for \textit{combined} data of $j$ and $\hat{i}$ until {$\rm{ACK_j}$} OR $r_T$}
\ELSE
\STATE{packet drop}
\ENDIF
\ENDFOR
\ENDFOR
%
%
%
\ENDIF
\ENDWHILE
\ENDIF
\end{algorithmic}
\end{algorithm}

Figure~\ref{fig:protocol_2} visualizes the procedure for two sender nodes, one with an impaired channel (node $i$) and the other with a good channel (node $j$), as an example.
The HARQ protocol at node $i$ decides on the bits to send in each round. If a NACK is received, the next portion of redundant bits will be transmitted. However, in our proposed algorithm, the impaired node is switched off after the first NACK arrives in round $1$. The collaborating neighbor $j$ overhears the impaired node, stores its data, and transmits it in round $2$.
Algorithm~\ref{algo:AHARQalgo} reports the pseudo-code executed by sender nodes in a neighborhood $\mathcal{N}_m$. For the neighbor discovery algorithm, an approach similar to the one used by the DIVE protocol~\cite{Petroccia16} can be employed. DIVE has a built-in mechanism to cope with unreliable channels. This approach can also be extended to share relevant link quality information when the network is deployed, thus supporting the cooperative strategy. This information can then be updated over time by piggybacking on regular data packets.

Data packets in the forward channel should be acknowledged successfully without error in the feedback transmission. The assumption of error-free feedback reception is not unreasonable since the length of this message is very short and therefore it can be protected by a strong channel coding technique. However, in a situation in which the ACK/NACK is lost, the timer expires to setup the retransmission process.
In the conventional scheme, if $T_i$ does not receive the ACK before a timeout expires, it will keep transmitting extra information in the next packets under the HARQ policy considering the previous channel state. However, in the proposed scheme, because of the collaboration among the nodes (i.e., $T_{\hat{c}}$), the probability of reception is increased by leveraging the statistical independency of the channels~(i.e., channel diversity).

\section{Performance Evaluation}\label{sec:Eval}

In this section, we provide the performance results when using data collected from the CMRE LOON testbed (Sect.~\ref{subsec:loon}) and during the REP18-Atlantic sea-trial (Sect.~\ref{subsec:atsea}). 
\begin{figure*}[t]
\centering
\begin{tabular}{cc}
\includegraphics[width=6.5cm]{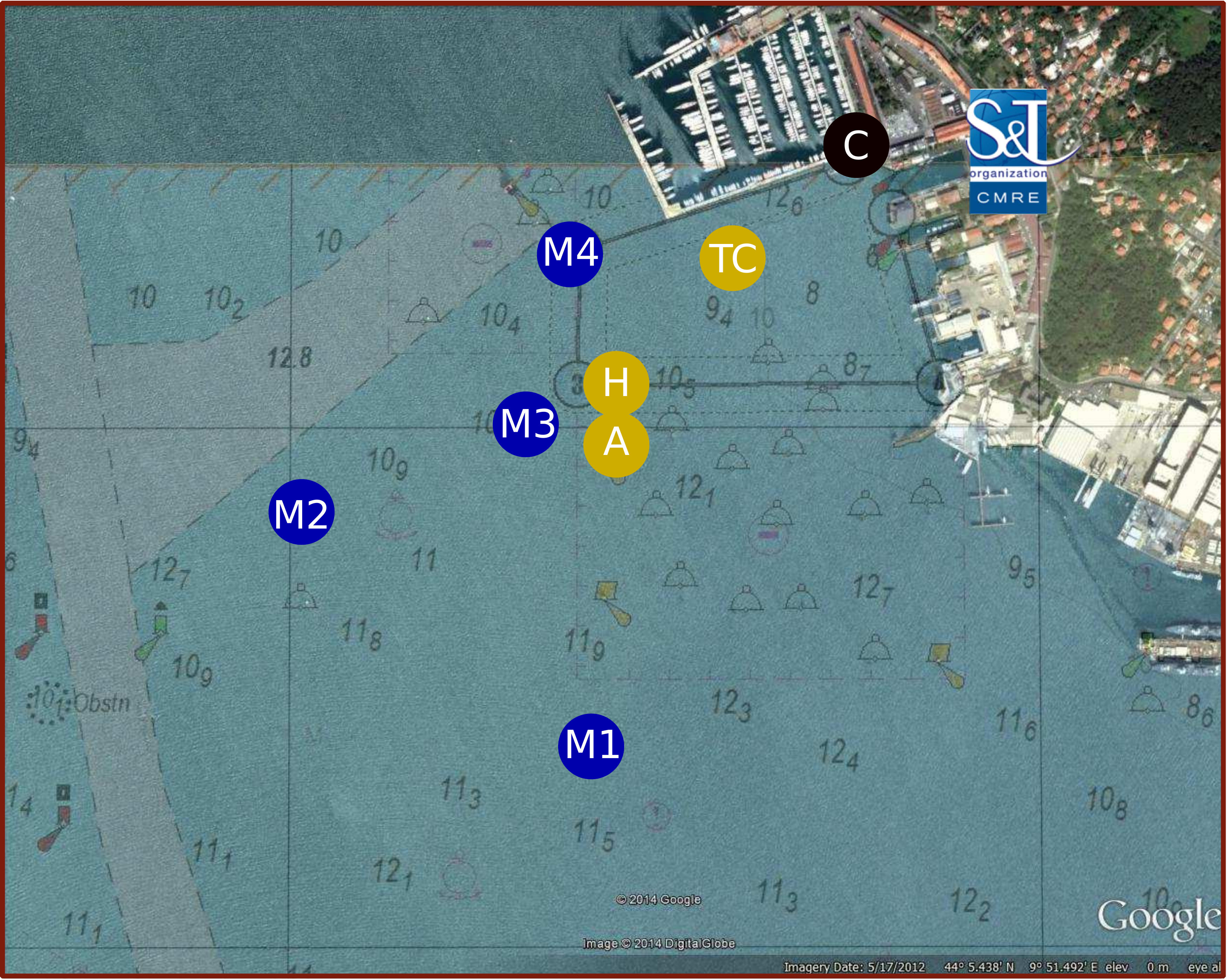}&
\includegraphics[width=7.35cm]{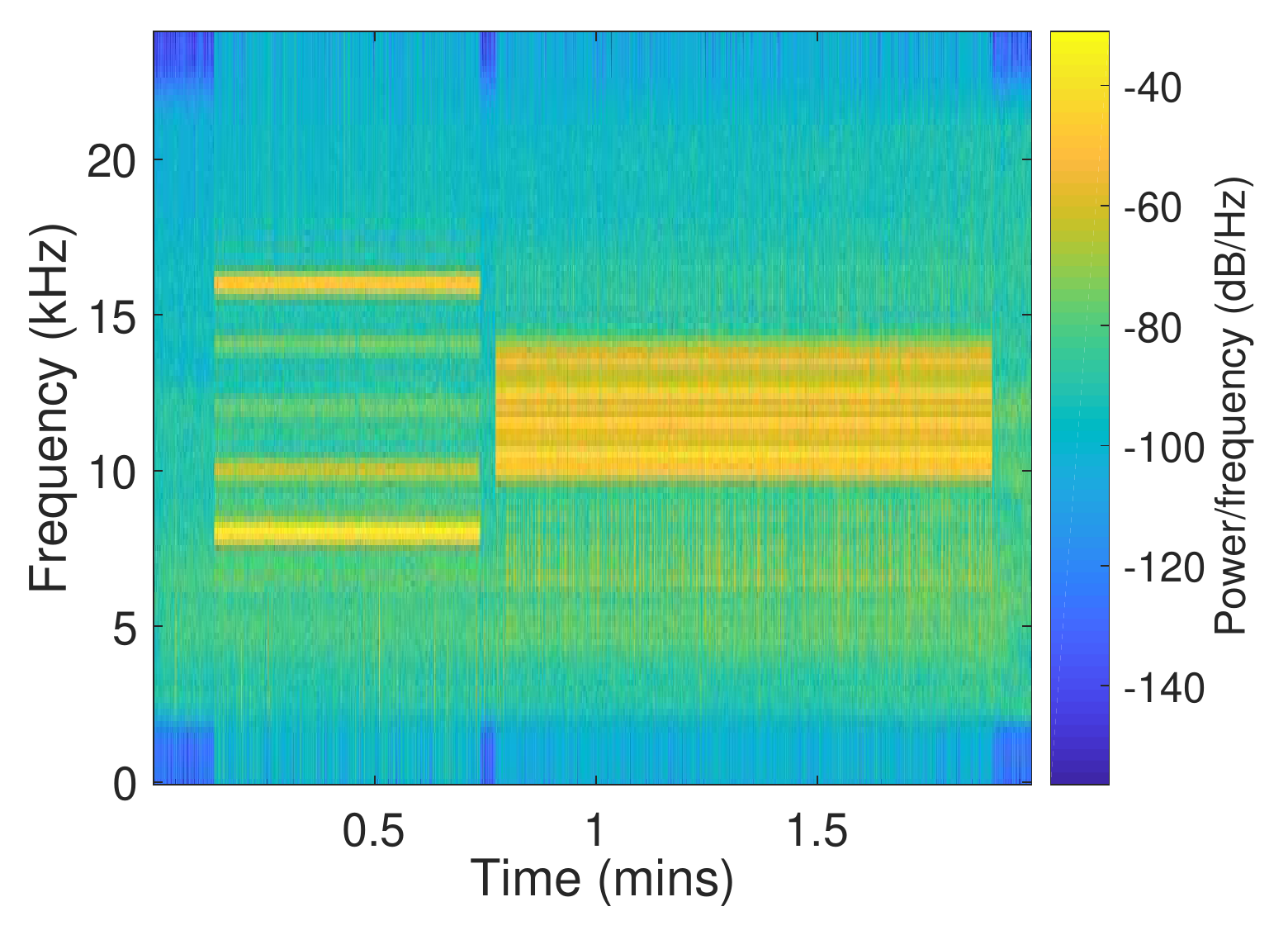}\\
\small (I) & \small (II) 
\end{tabular}
\caption{\small (I)~Geographic configuration of the CMRE LOON testbed, in the Gulf of La Spezia, Italy, where $M1$ to $M4$ are the modem tripods, $C$ is the shore side container lab where the control station is located, $TC$ is the thermistor chain, $H$ is the hydrophone array, and $A$ is an ADCP. 
(II)~Spectrum of a sample received signal from the LOON in two successive transmission time slots while the spreading length and coding rate have changed.}\label{fig:mapANDspect}
\end{figure*}

\subsection{CMRE LOON Testbed}\label{subsec:loon}

\textbf{Testbed:}
The geographic configuration of the CMRE LOON testbed is depicted in Fig.~\ref{fig:mapANDspect}(I) for underwater communications and networking. It consists of four bottom-mounted tripods~(M1-M4) installed at a depth of about $10~\rm{m}$. Each tripod is equipped with heterogeneous communications technologies and sensors, and it is cabled to a shore control station~(C) providing data connection and power supply. 
The LOON tripods also support arbitrary waveform transmission/recording. 
Additionally, the LOON includes a high-definition acoustic data acquisition system (at frequencies above $1~\rm{kHz}$) from an array of hydrophones~(H), a thermistor chain~(TC), sound velocity sensors, an Acoustic Doppler Current Profiler~(ADCP) with waves measurement~(A), and a meteorological station. These sensors are used to correlate the characteristics of the acoustic channel with the performance of the investigated protocols. The LOON provides therefore a comprehensive data set of environmental, acoustic, and packet measurements to study the communication processes at different communication layers.

\textbf{Experiment Settings:}
A variety of scenarios can be considered in the shallow-water environment where the LOON is deployed to capture the system outputs.  
We considered a point-to-point transmission from node $M4$ to $H$ for modeling the link of several rounds of transmissions. Packets are transmitted using baseband Binary Phase Shift Keying~(BPSK) and Quadrature Phase Shift Keying~(QPSK) modulations over the passband channel $4-19~\rm{kHz}$ 
by exploiting Reed-Solomon channel coding, $(7,3)$ or $(15,9)$. A logistic map is used to generate a chaotic spreading code with various lengths, i.e., $SL=[10,40]$. As an example, we have measured the average Signal-to-Noise Ratio~(SNR) equal to $32.68~\rm{dB}$ for a QPSK transmitted signal with $SL=22$, and an average Bit Error Rate~(BER) of approximately $4.6365 \times 10^{-4}$ is achieved.

\begin{figure*}[!t]
\centering
\begin{tabular}{cc}
\hspace{-0.2in}
\includegraphics[width=8.4 cm]{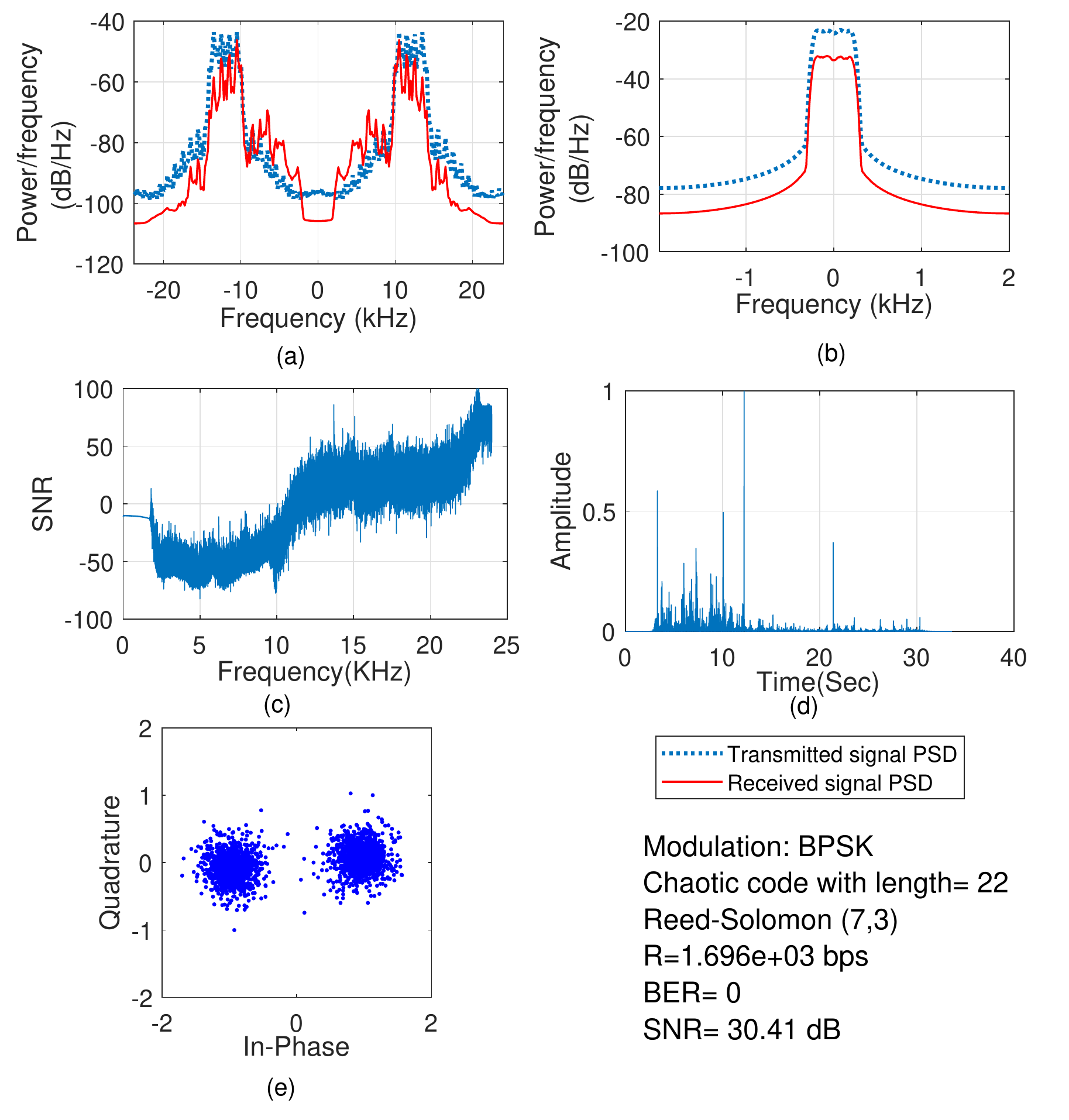} &
\includegraphics[width=8.4 cm]{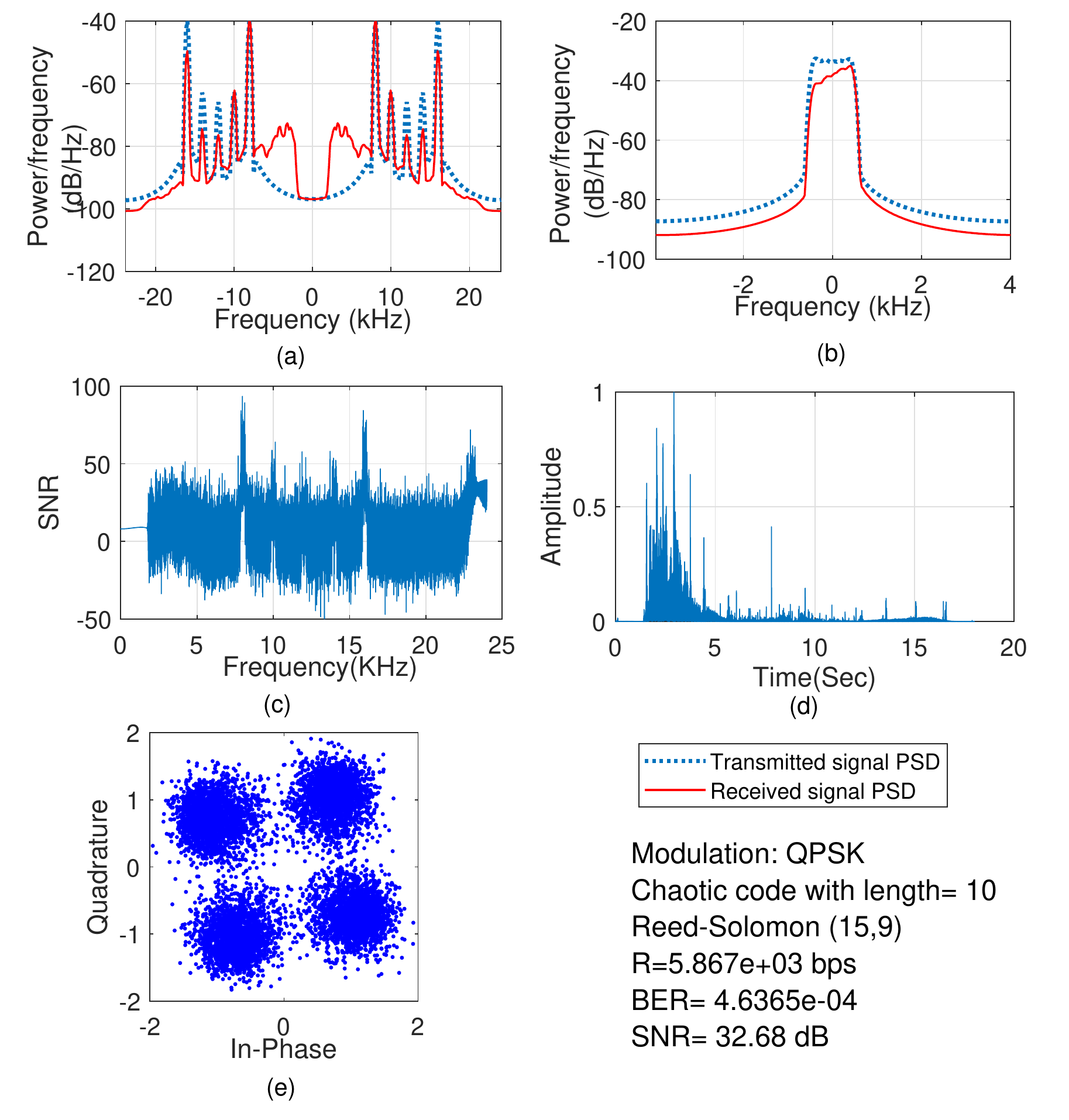} \vspace{-0.7 cm}\\
\small (I) & \small (II) 
\end{tabular}
\caption{\small (I)~and (II) show two experiments with different parameters; (a)~Power Spectral Density~(PSD) of the transmitted and received passband signals; (b)~PSD of the baseband decoded received signal in comparison with transmitted signal; (c)~Signal-to-Noise Ratio~(SNR) of the received signal per frequency; (d)~Experienced channel profile; (e)~Constellation of the equalized baseband received signal.}
\label{fig:per1}
\end{figure*}
\begin{figure*}
\centering
\begin{tabular}{ccc}
\hspace{-0.6in}
\includegraphics[width=6.5cm
]{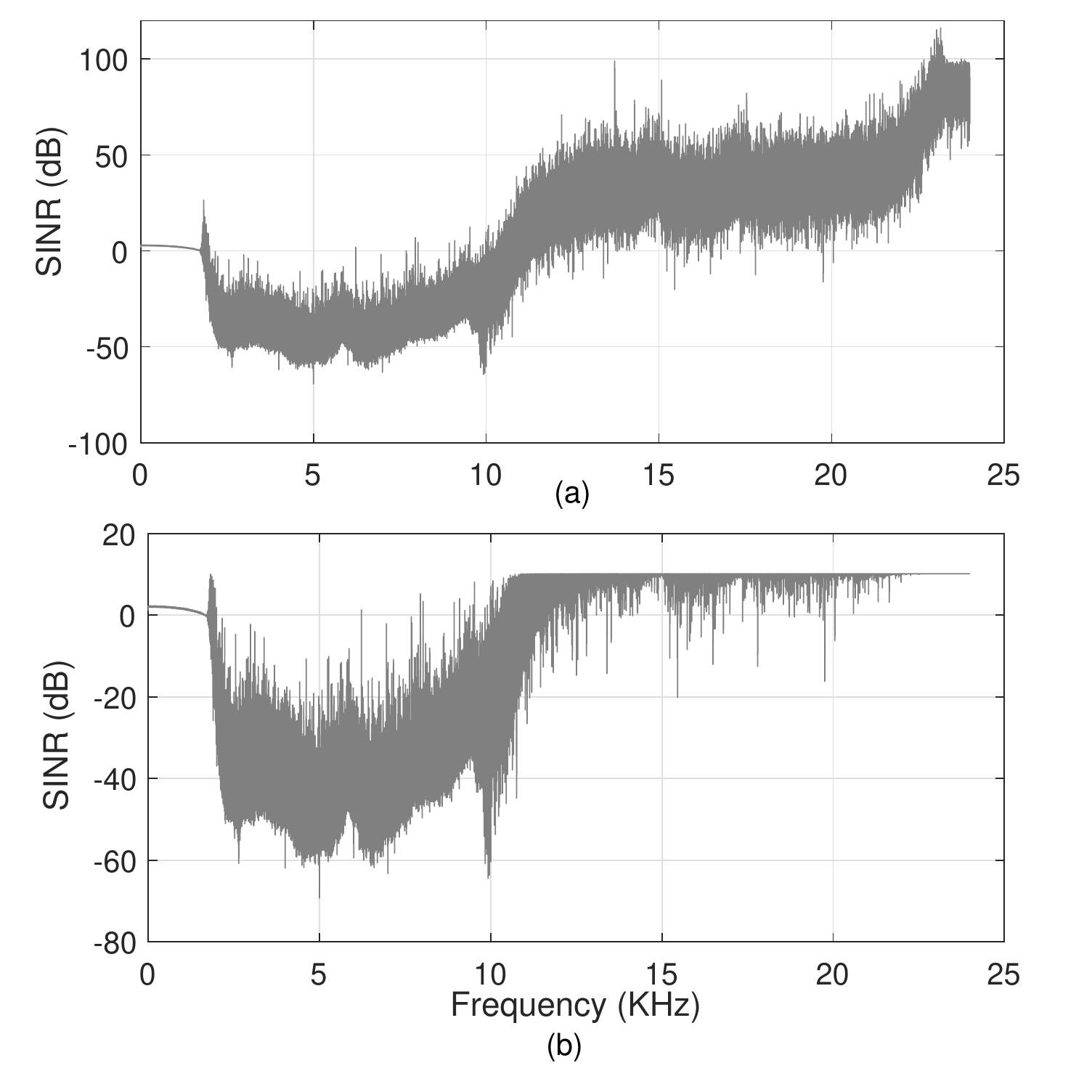} &
\hspace{-8mm}
\includegraphics[width=6.5cm
]{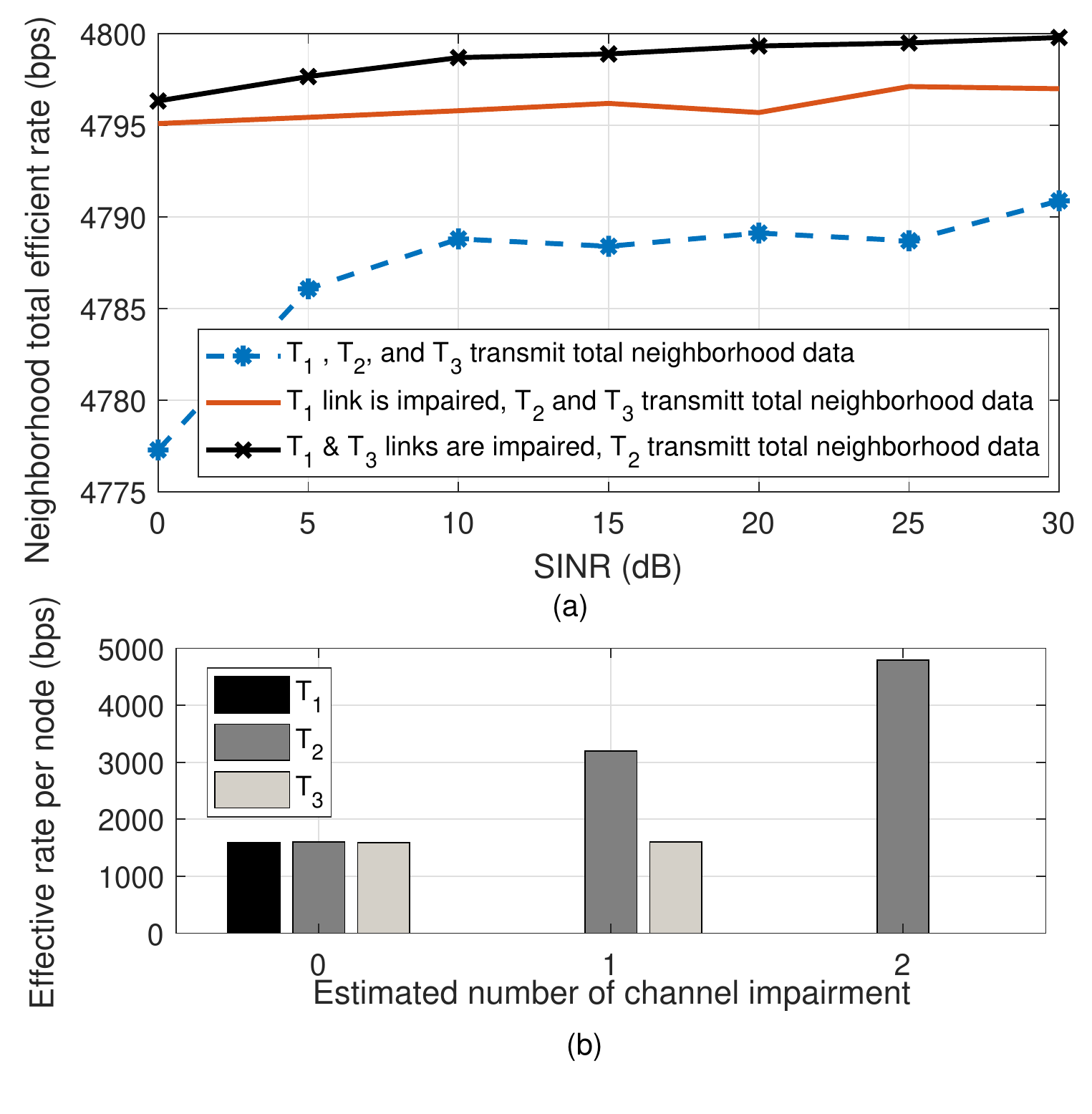} &
\hspace{-7mm}
\includegraphics[width=6.85cm
]{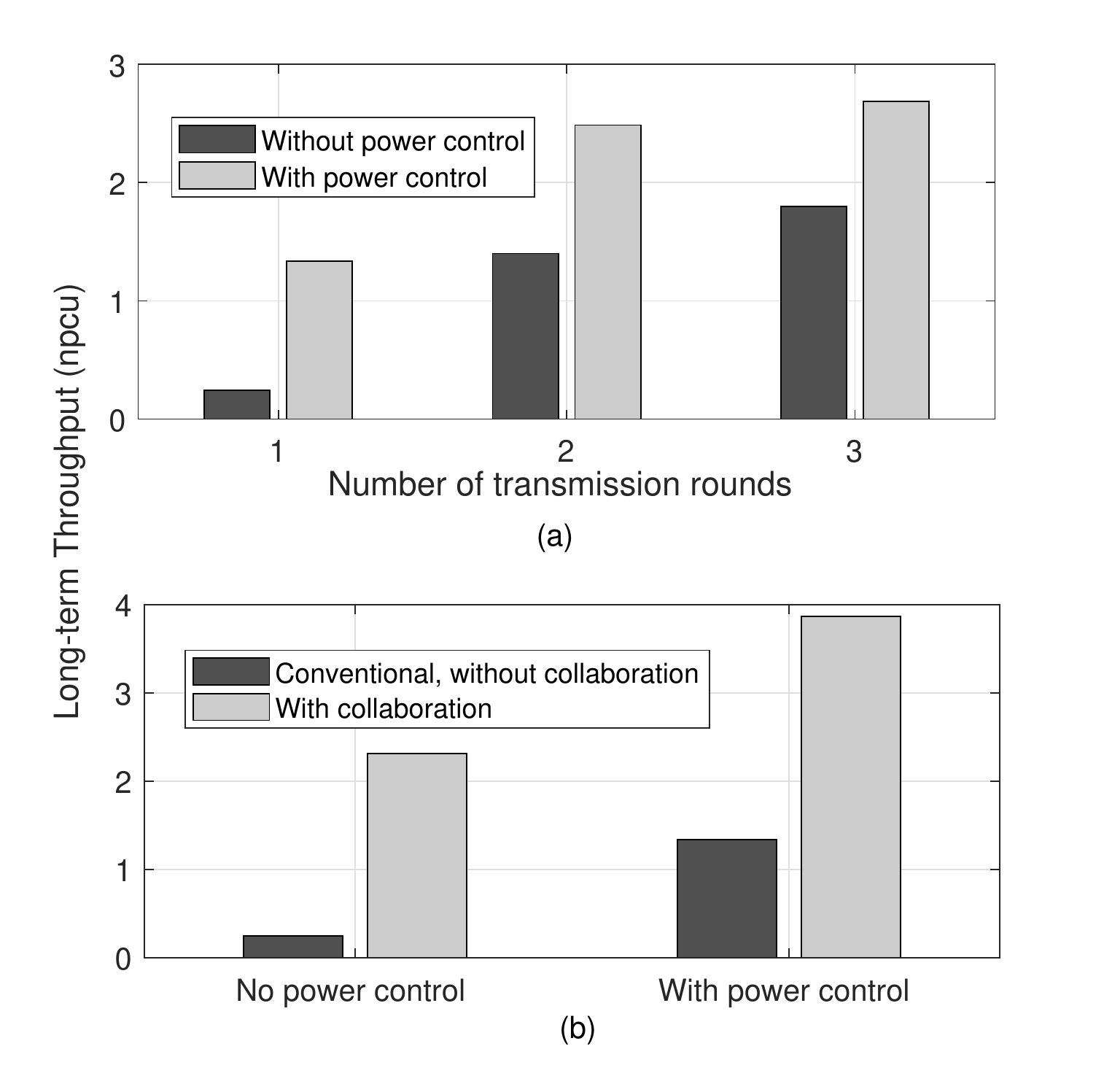} \vspace{-0.35cm}\\
\vspace{-0.2 cm}
\small (I) & \small (II)  & \small (III)
\end{tabular}
\caption{\small (I)~SINR (simulation) when the number of transmitting nodes in a neighborhood is (a)~one; (b)~three. (II)~(a)~Efficient rate of the neighborhood; (b)~Effective rate per transmitting node with channel impairments. (III)~Long-term throughput in nats-per-channel-use~(npcu) (a)~for different maximum number of transmission rounds when no channel impairment occurs and with/without power control; (b)~comparing traditional HARQ with our collaborative method, with/without power control.}
\vspace{-3mm}
\label{fig:sim_results}
\end{figure*}

\textbf{Simulation Settings:}
We focus on the collaboration among transmitting nodes and assume that ACK/NACK feedback links are free of errors. Algorithm~\ref{algo:AHARQalgo} has however a mechanism with a timer to retransmit the data if a feedback is not received within the expected time. Nodes in the same neighborhood can overhear each other, while nodes in adjacent areas do not receive the data since the chaotic CDMA sequence protects from unauthorized overhearing. Simulation are conducted for a neighborhood of $3$ nodes. We use the data collected using the LOON testbed to model a multiuser scenario, then we optimize the parameters, as described in Sect.~\ref{sec:Probsol}. The computed values are passed through the channels extracted from the LOON in a close-loop manner.  
We evaluate the system performance in MATLAB by considering the following metrics: SINR, long-term throughput ($\eta$), neighborhood efficiency rate, and effective rate per node.

\textbf{Results:} 
Figure~\ref{fig:mapANDspect}(II) shows the frequency spectrum of a sample received signal from the LOON while the spreading length and coding rate changes for two successive transmitted signals. In Figs.~\ref{fig:per1}(I) and \ref{fig:per1}(II), two experiments with different settings are shown (for BPSK and QPSK scenarios, respectively). The PSD of the transmitted and received signals in passband and decoded baseband are plotted for comparison. Received SNR versus bandwidth, channel profile for the duration of the transmission, and scatter plot of the estimated symbols are provided. The transmitted signal parameters, BER, and SNR are also included in the figure. In Fig.~\ref{fig:sim_results}(I), the received SINR in a neighborhood of three nodes is presented to investigate the effect of multiuser interference. Figure~\ref{fig:sim_results}(I-a) presents the case where only one node in the area is transmitting. In this case, without interference, the received signal has a considerably better SINR. In Fig.~\ref{fig:sim_results}(I-b), the data transmission in the area is performed by three nodes, so there is a multi-user interference. Figure~\ref{fig:sim_results}(II-a) depicts the total efficient rate in a neighborhood of three nodes.  
The plot shows how the collaboration strategy handles channel impairments and distributes the traffic load in the neighborhood. As the result of collaboration, when there are fewer nodes to perform data transmission, multiuser interference drops and spectral efficiency improves. Figure~\ref{fig:sim_results}(II-b) presents the effective received rate per node. The plot shows that in the case where an impairment occurs in $T_1$ link, $T_2$ collaborates in data transmission. In Fig.~\ref{fig:sim_results}(III), long-term throughput for the proposed collaborative method is investigated. In Fig.~\ref{fig:sim_results}(III-a), $\eta$ is plotted for different values of $r_T$ when none of the nodes experiences channel impairment. The plot confirms that power control can improve the long-term throughput. Finally, in Fig.~\ref{fig:sim_results}(III-b), collaborative HARQ is compared with the conventional method to confirm that collaboration improves long-term throughput under channel impairment. The figure also shows the positive effect of power control. 

\begin{figure*}[t]
\centering
\begin{tabular}{cc}
\includegraphics[width=7cm
]{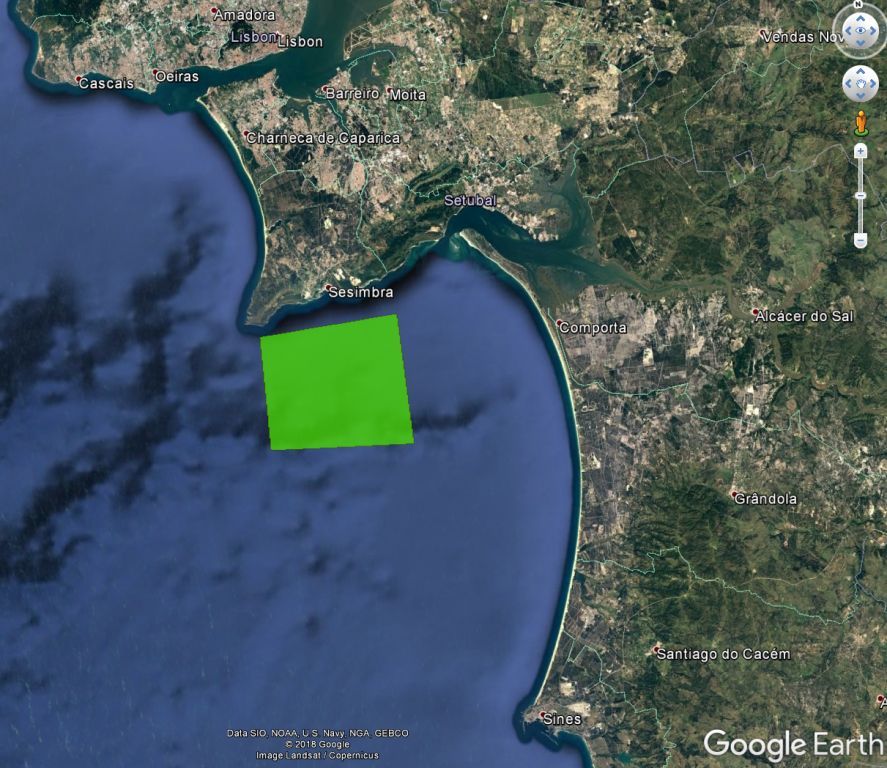}&
\includegraphics[width=8.8cm
]{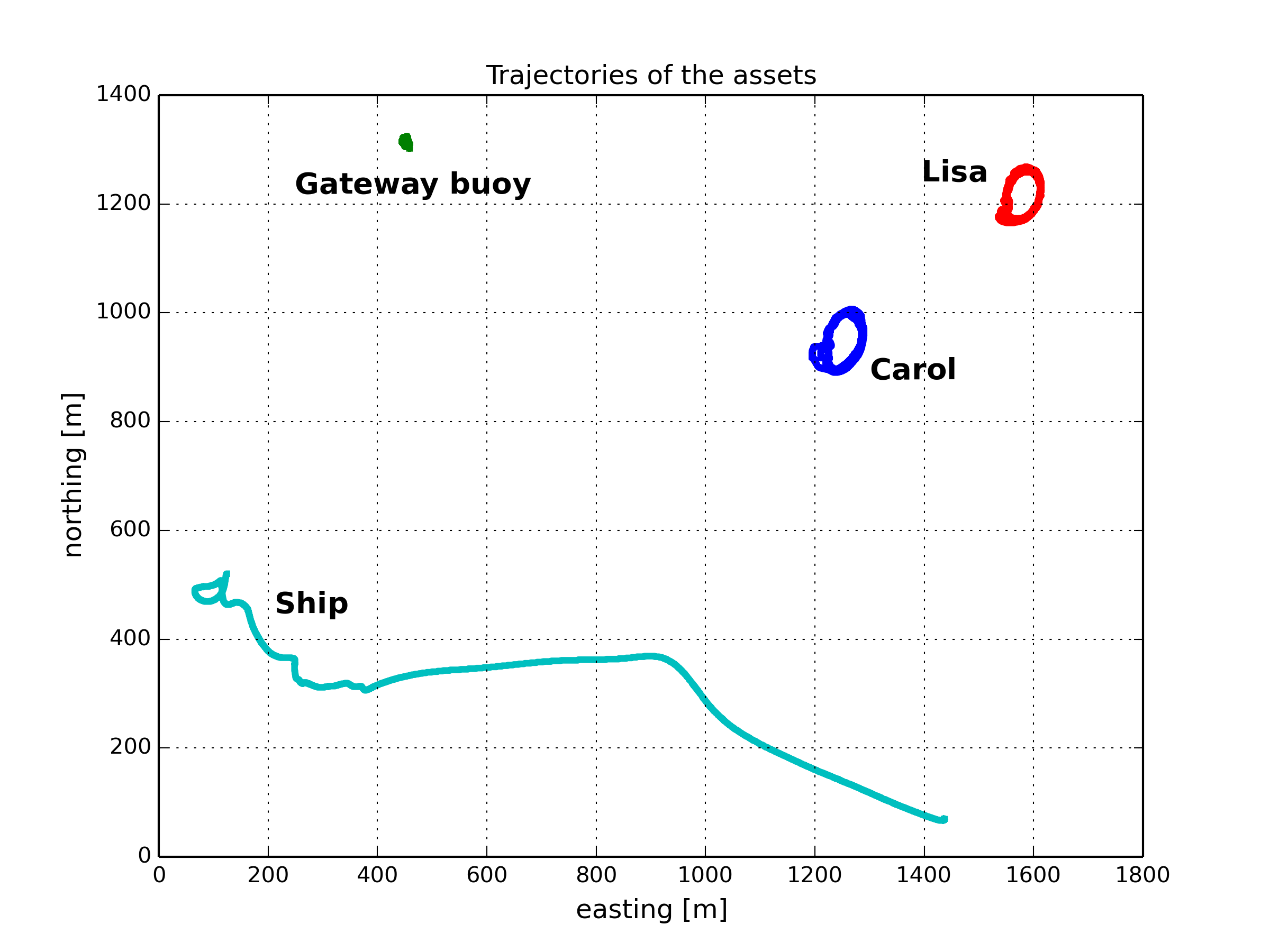}\\
\small (I) & \small (II) 
\end{tabular}
\caption{\small (I)~REP18 area of operations; 
(II)~Trajectories of the nodes.}\label{fig:rep18area_nodenav}
\end{figure*}

\subsection{Sea Experiment}\label{subsec:atsea}

\textbf{Experiment Setup:}
Sea experiments were conducted during the REP18-Atlantic (Recognize Environmental Picture) trial organized by CMRE, the Portuguese Navy~(PRT-N), and the Faculty of Engineering of the University of Porto~(FEUP). The trial took place place from the $1^{st}$ to the $20^{th}$ of September 2018, in the Atlantic Ocean off the coast of Portugal between Sines and Sesimbra. The area of operations is depicted in Fig.~\ref{fig:rep18area_nodenav}(I). The scope of this exercise was to investigate, evaluate, and demonstrate novel technologies and solutions in the domain of underwater communications and networking; as well as aerial, surface, and underwater robotic solutions and autonomous strategies.

\begin{table}[!htbp]
\centering
\caption{Deployed CMRE assets.}
\small
\begin{tabular}{c m{
12cm
}}
\hline
\\
  \begin{minipage}{.18\textwidth} 	\includegraphics[width=\linewidth, height=30mm]{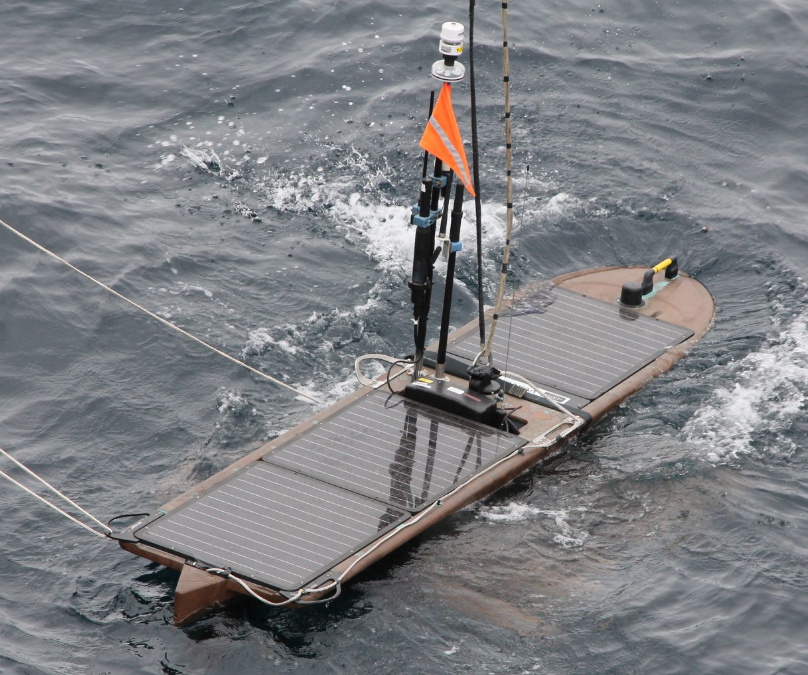} \end{minipage} 
&  \textbf{WaveGlider SV3} (x2): the wave glider is a self-propelled unmanned surface vehicle which uses wave motion to navigate. The SV3 version is also equipped with an auxiliary propeller. The wave glider enables long duration exploration and monitoring operations. Two wave gliders (named Lisa and Carol) were deployed, both equipped with an embedded board to run locally the required software and the capability to record acoustic signals using the iclisten smart hydrophone~\cite{iclisten}. Lisa was also equipped with the capability of transmitting arbitrary waveforms using the Neptun~T313 transducer~\cite{NeptunT313}; On Lisa the acoustic payload was deployed at a depth of $\sim40$m, while for Carol the depth was $\sim20$m.\\
 
 \begin{minipage}{.18\textwidth} 	\includegraphics[width=\linewidth, height=37mm]{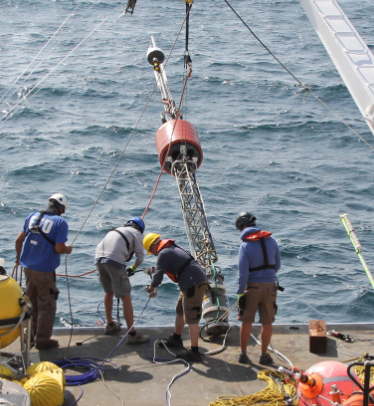} \end{minipage}
&  \textbf{Moored gateway buoy} (x1): Moored buoy equipped with dual radio connectivity (Wi-Fi $2.4$~GHz and Freewave $900$~MHz), an embedded board to run locally the required software and the capability to transmit/receive arbitrary waveforms using the Neptun~T313 transducer and the iclisten smart hydrophone, respectively. The acoustic payload was deployed at a depth of $\sim80$m.\\
 
 \begin{minipage}{.18\textwidth} 	\includegraphics[width=\linewidth, height=30mm]{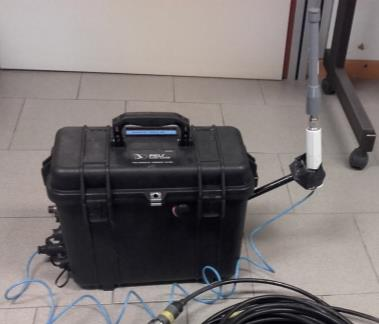} \end{minipage} 
&  \textbf{Manta portable node} (x1): this is a portable node including radio connectivity (Wi-Fi $2.4$~GHz), an embedded board to run locally the required software, and the capability to transmit/receive arbitrary waveforms using the ITC3013~\cite{ITC3013} transducer and the  iclisten smart hydrophone, respectively. It was deployed from the NRP Almirante Gago Coutinho ship during the conducted activities, with the acoustic payload at a depth of $\sim20$m. The ship was left drifting during the experiment to avoid impacting the data collection with the noise produced by the propellers.\\
\hline
\end{tabular}
\label{table:sea_assets}
\end{table}

During the REP18 sea trial, dedicated tests were conducted to investigate the use CDMA signals for underwater acoustic networking. These tests were scheduled in the night between the $8^{th}$ and the $9^{th}$ of September making use of four CMRE assets. In the area of the experiment a maximum depth of $\sim130~\rm{m}$ was experienced. Table~\ref{table:sea_assets} details about the deployed assets, while Fig.~\ref{fig:rep18area_nodenav}(II) displays about the nodes trajectories and distances during the conducted experiment. A source level of $184~dB~re$~$\mu$Pa@1m was considered at the transmitter, which is in line with that of many commercial acoustic modems currently available on the market.

\textbf{Experiment Description:}
Two main scenarios were considered, i.e., single transmitter and simultaneous synchronized transmitters. Four nodes were deployed, as depicted in Fig.~\ref{fig:rep18area_nodenav}(II). One node, named Carol, was equipped with reception capability. The remaining three nodes (i.e., Gateway, Lisa and Manta/Ship) were instead provided with both transmission and reception capability. Multiple signals were transmitted and received during the conducted experiments.
The usable bandwidth was between $7-16~\rm{kHz}$ and the receivers were able to record the sound with the rate of $64~\rm{kHz}$. The modulation methods were BPSK and QPSK. A chaotic DS-CDMA sequence with Logistic and Bernoulli maps were created with the spreading lengths of $10$, $30$, $40$. A Reed-Solomon channel coding was used with different strengths as $(7,3)$ $(15,9)$, $(31,19)$, $(31,21)$, $(31,23)$, $(31,25)$.   

\begin{figure*}
\centering
\begin{tabular}{ccc}
\hspace{-0.45in}
\includegraphics[width=6cm
]{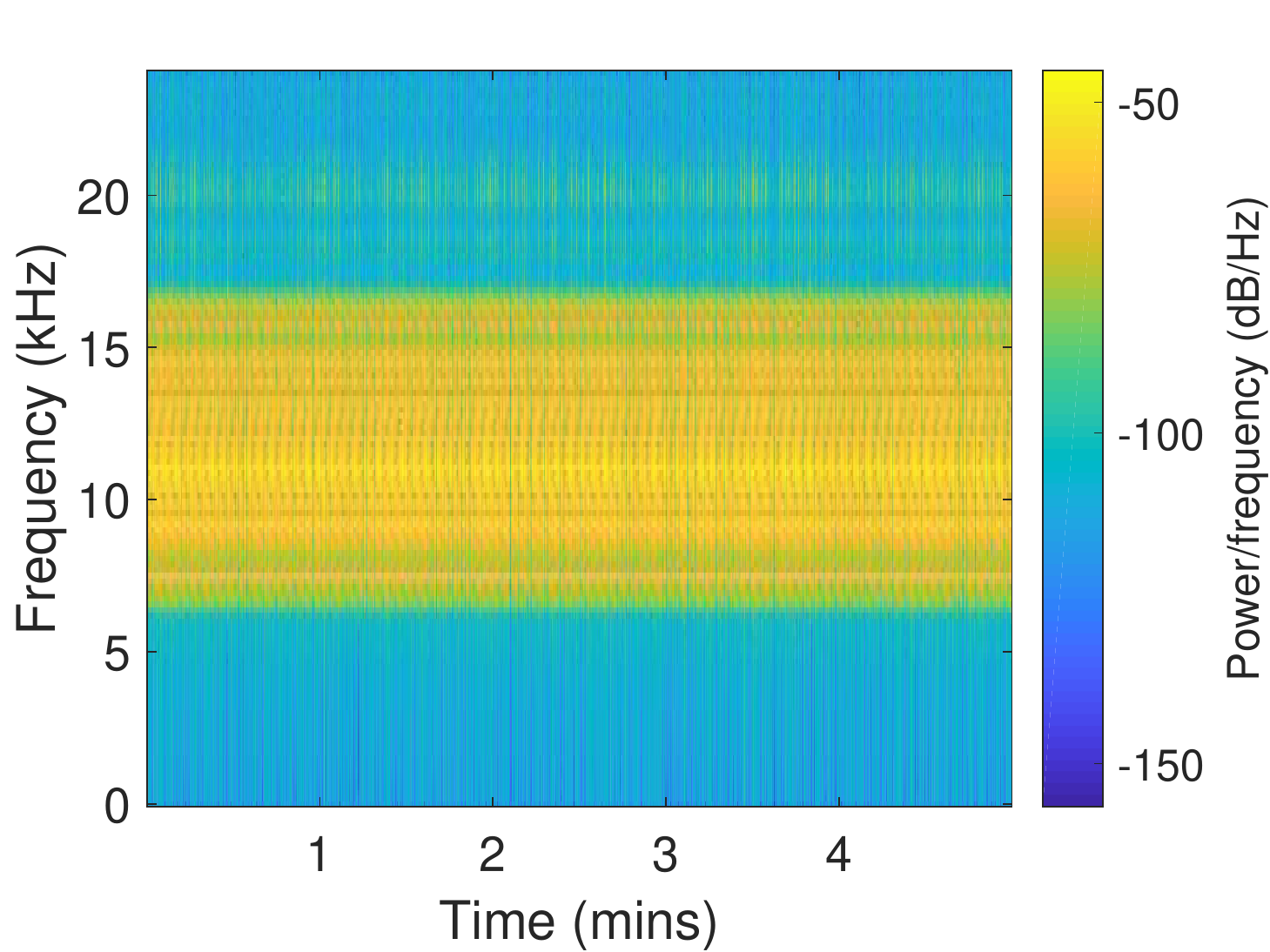} &
\hspace{-5mm}
\includegraphics[width=6cm
]{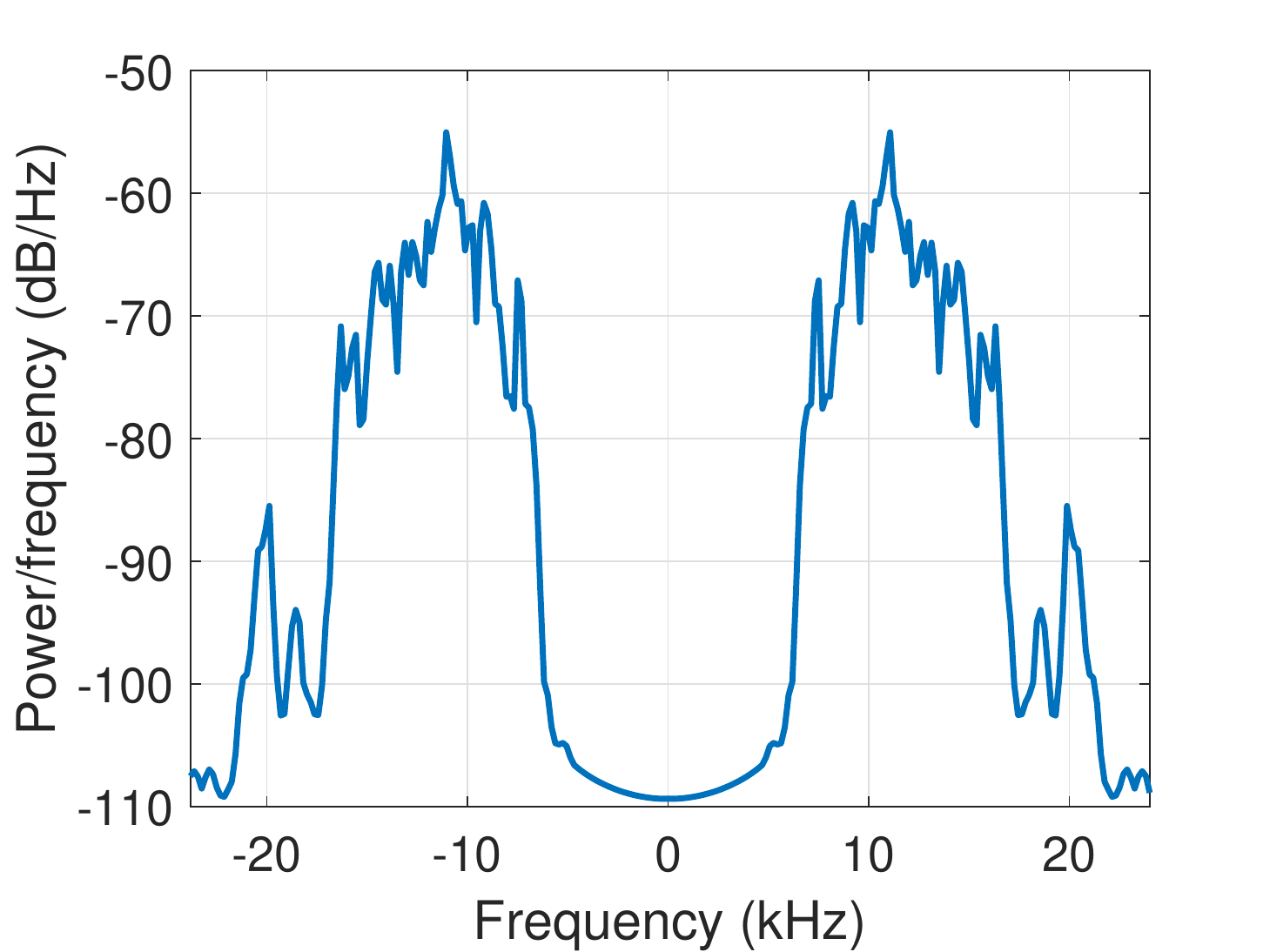} &
\hspace{-11mm}
\includegraphics[width=6.6cm
]{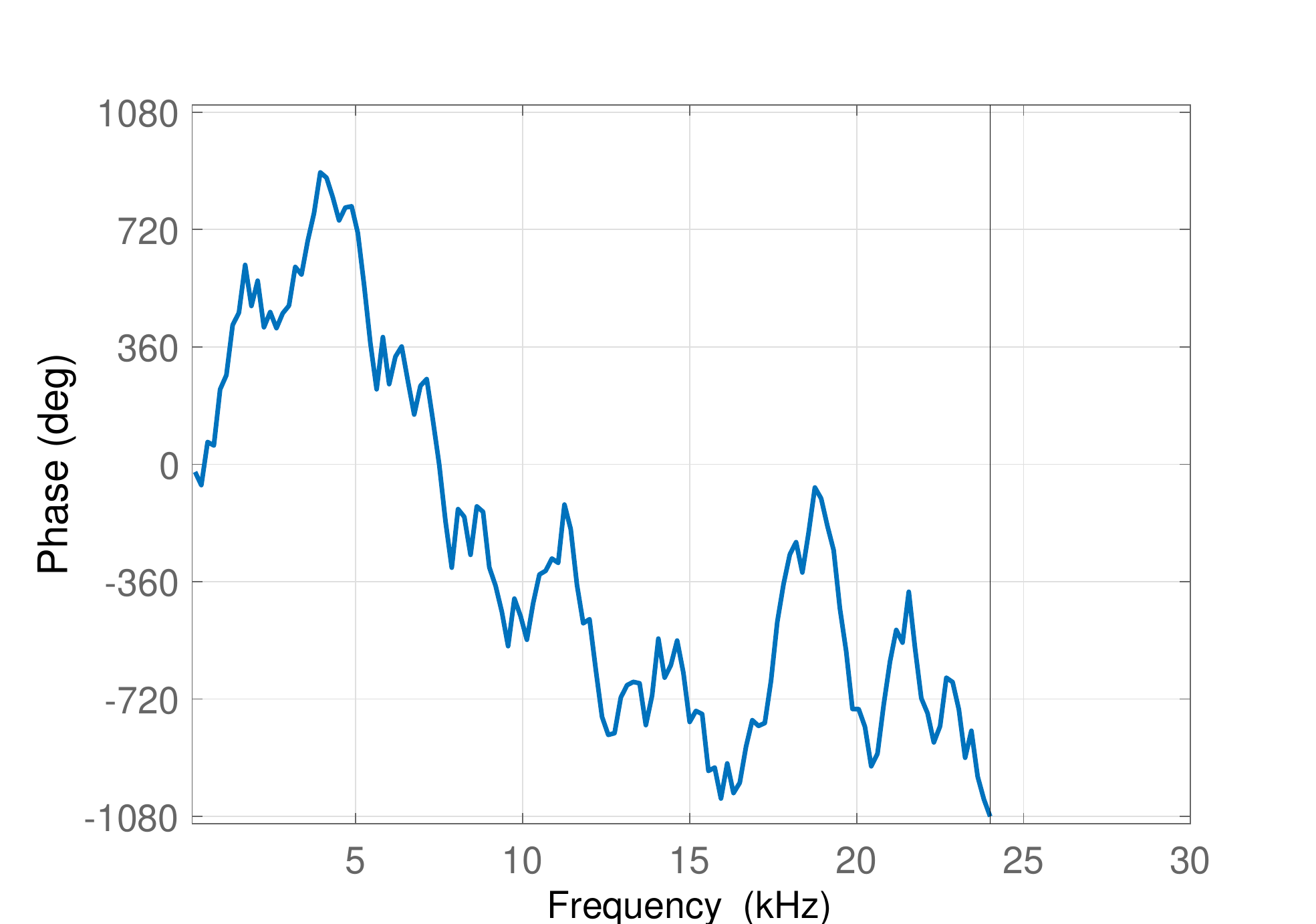} \\
\vspace{-1mm}
\small (a) &  \small (a)  & \small (a)\\
\hspace{-0.45in}
\includegraphics[width=6cm
]{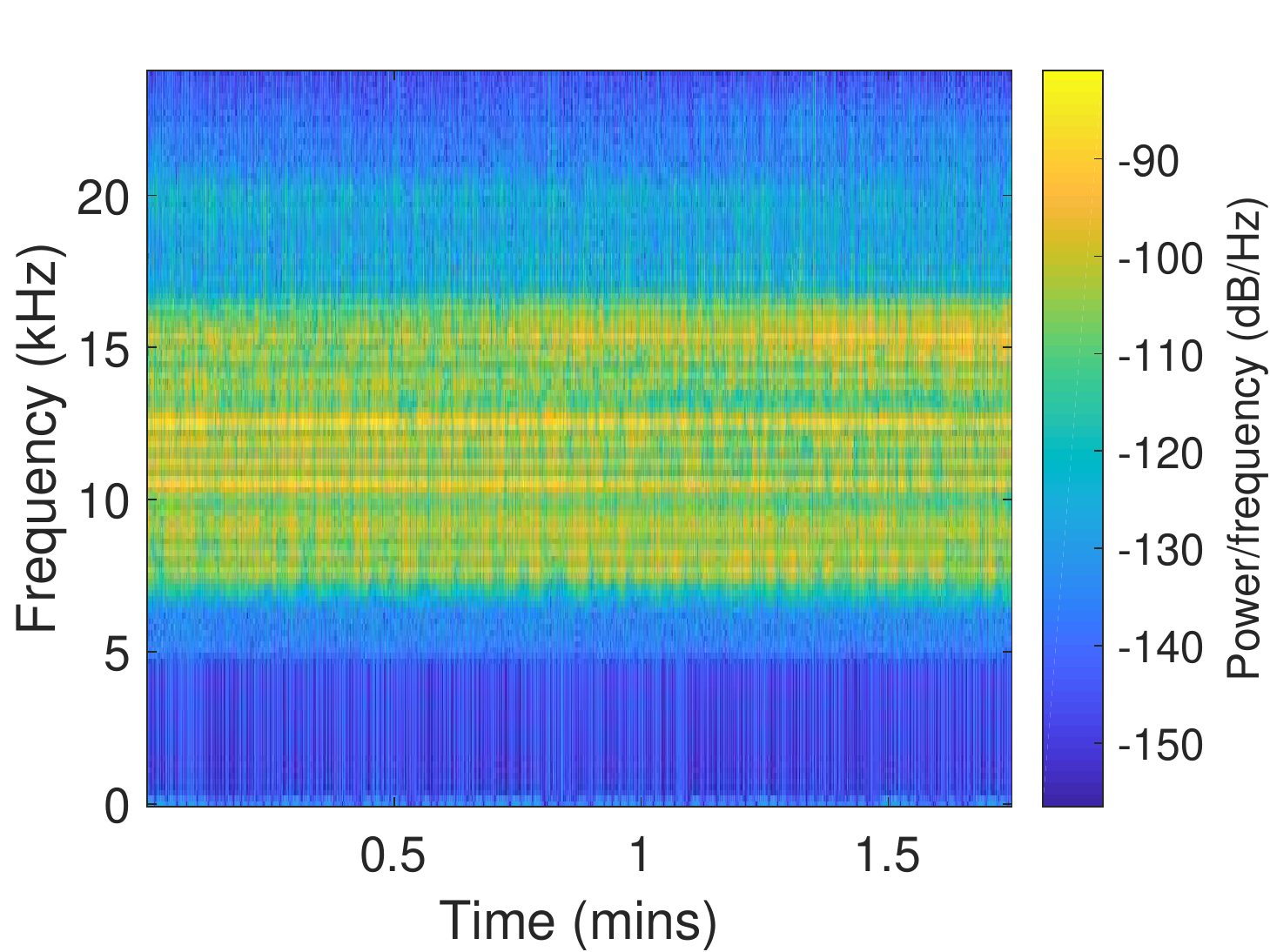} &
\hspace{-5mm}
\includegraphics[width=6cm
]{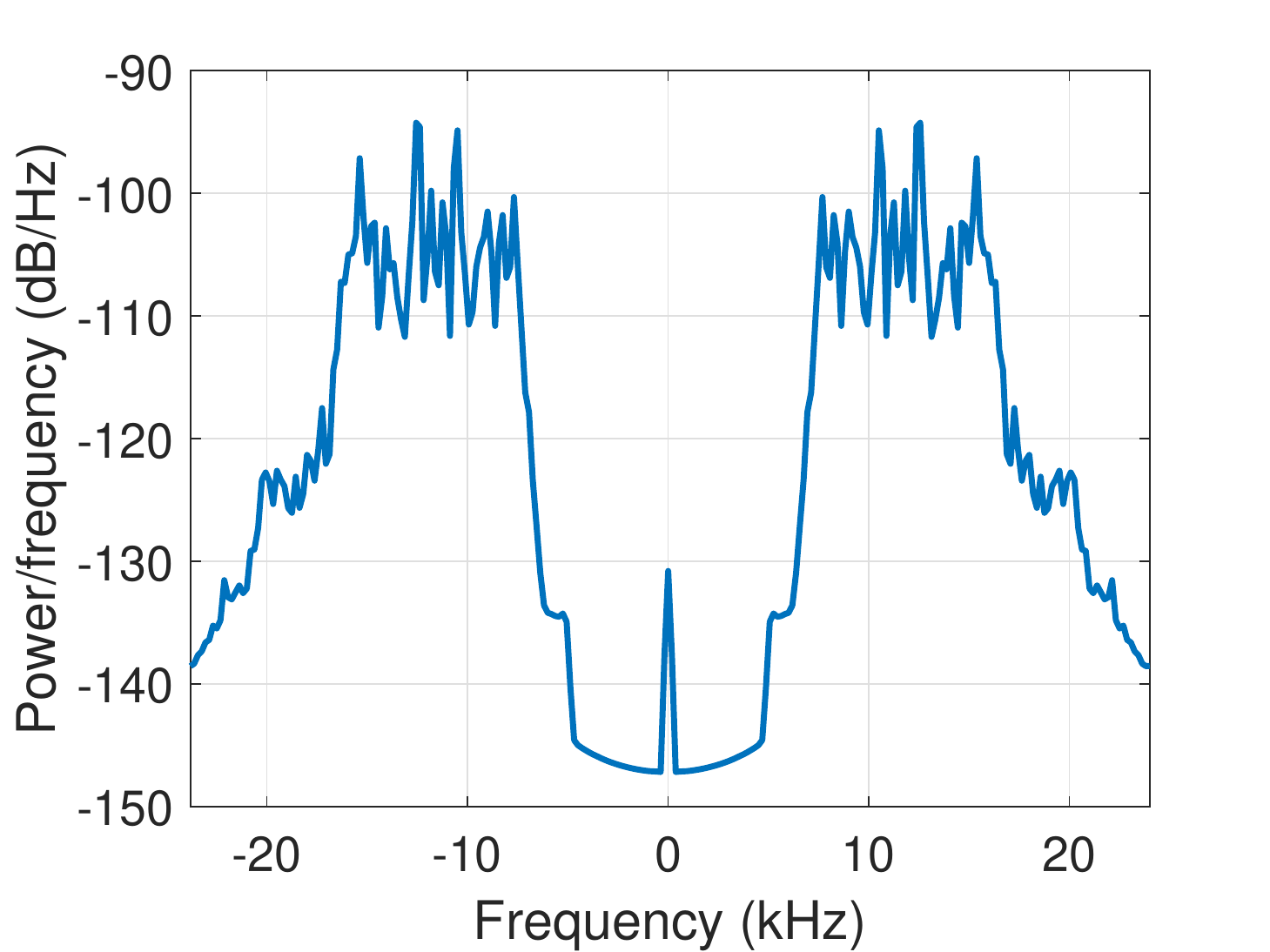} &
\hspace{-11mm}
\includegraphics[width=6.6cm
]{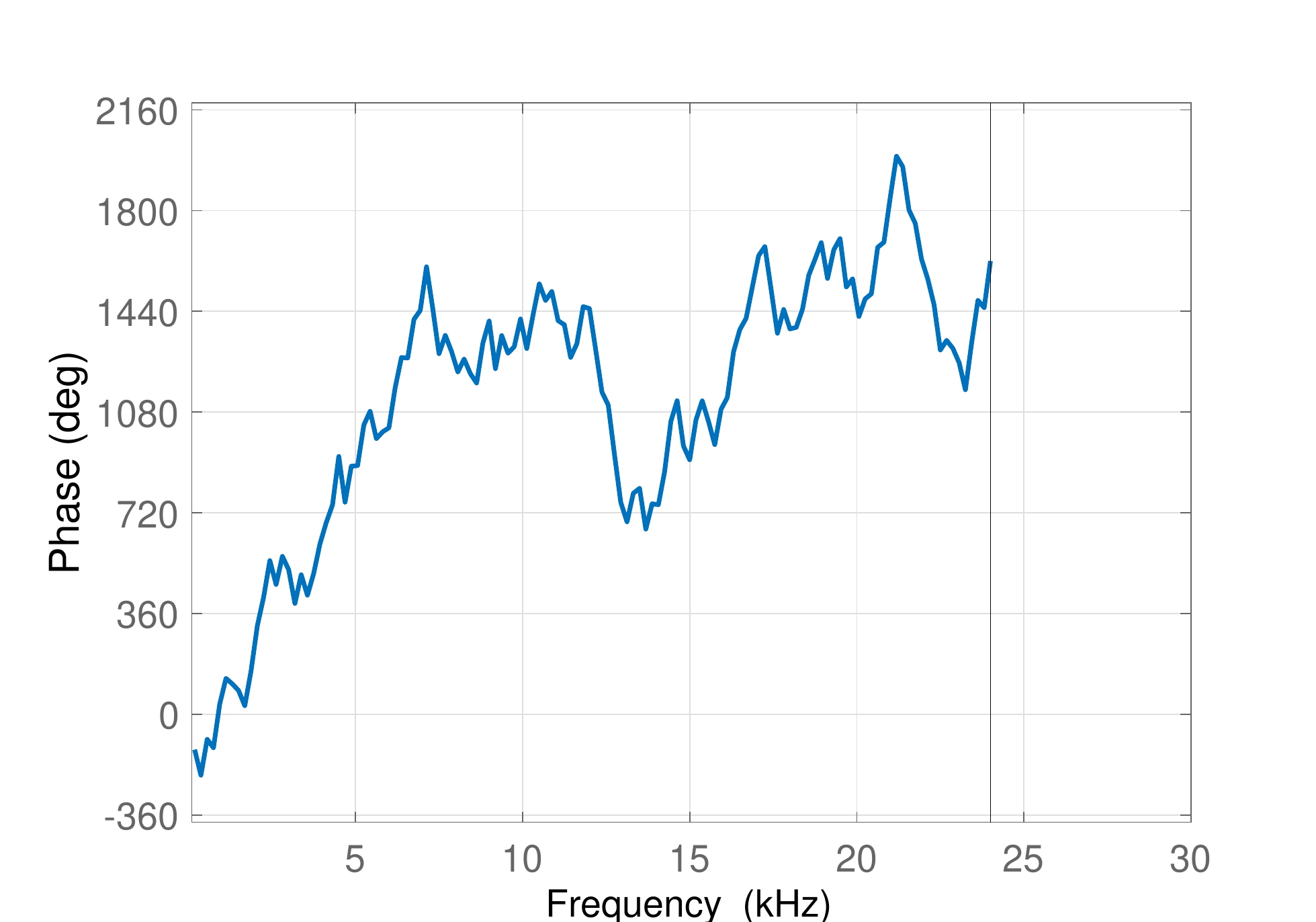} \\
\vspace{-1mm}
\small (b) &  \small (b)  & \small (b)\\
\hspace{-0.45in}
\includegraphics[width=6cm
]{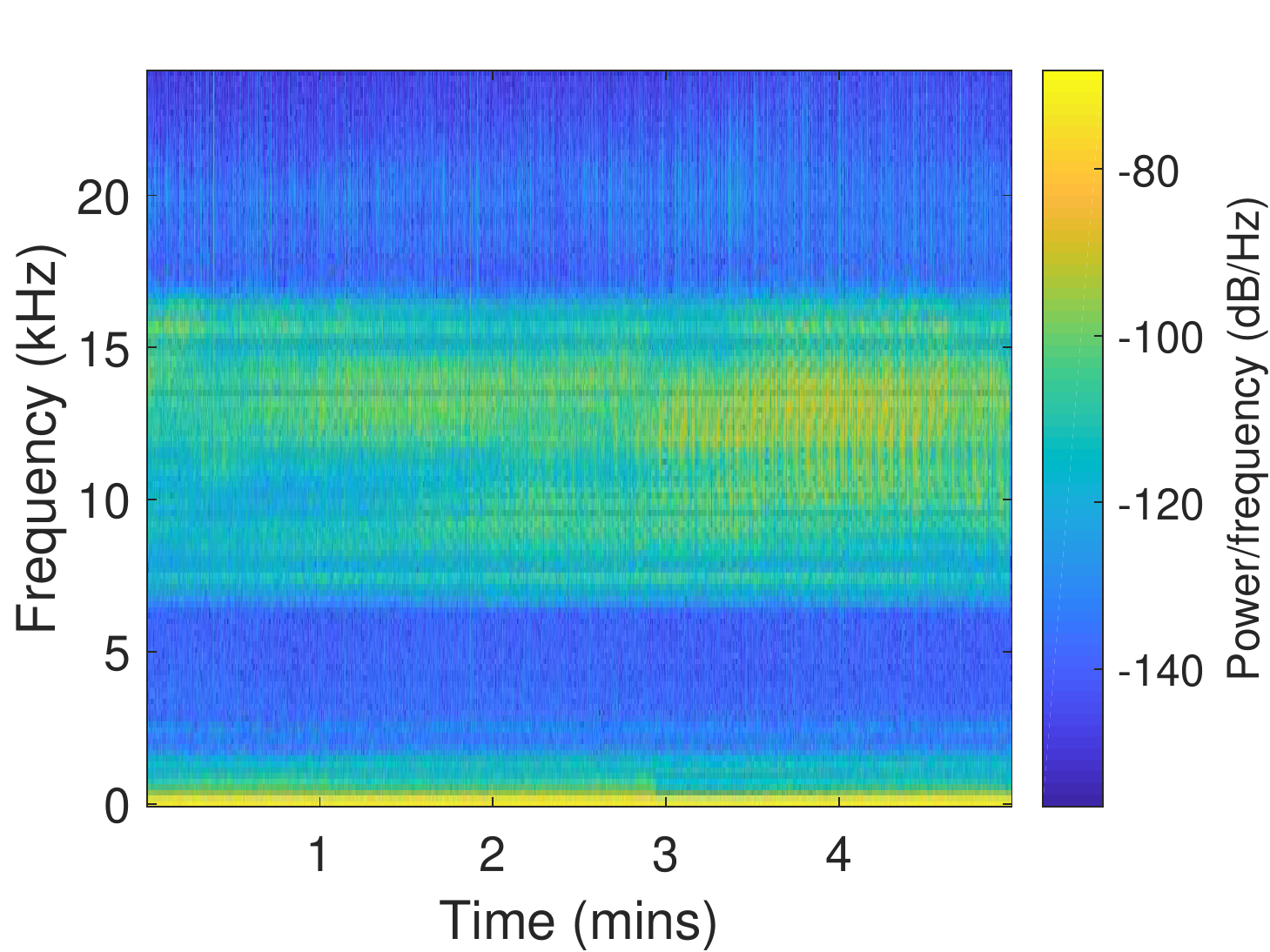} &
\hspace{-5mm}
\includegraphics[width=6cm
]{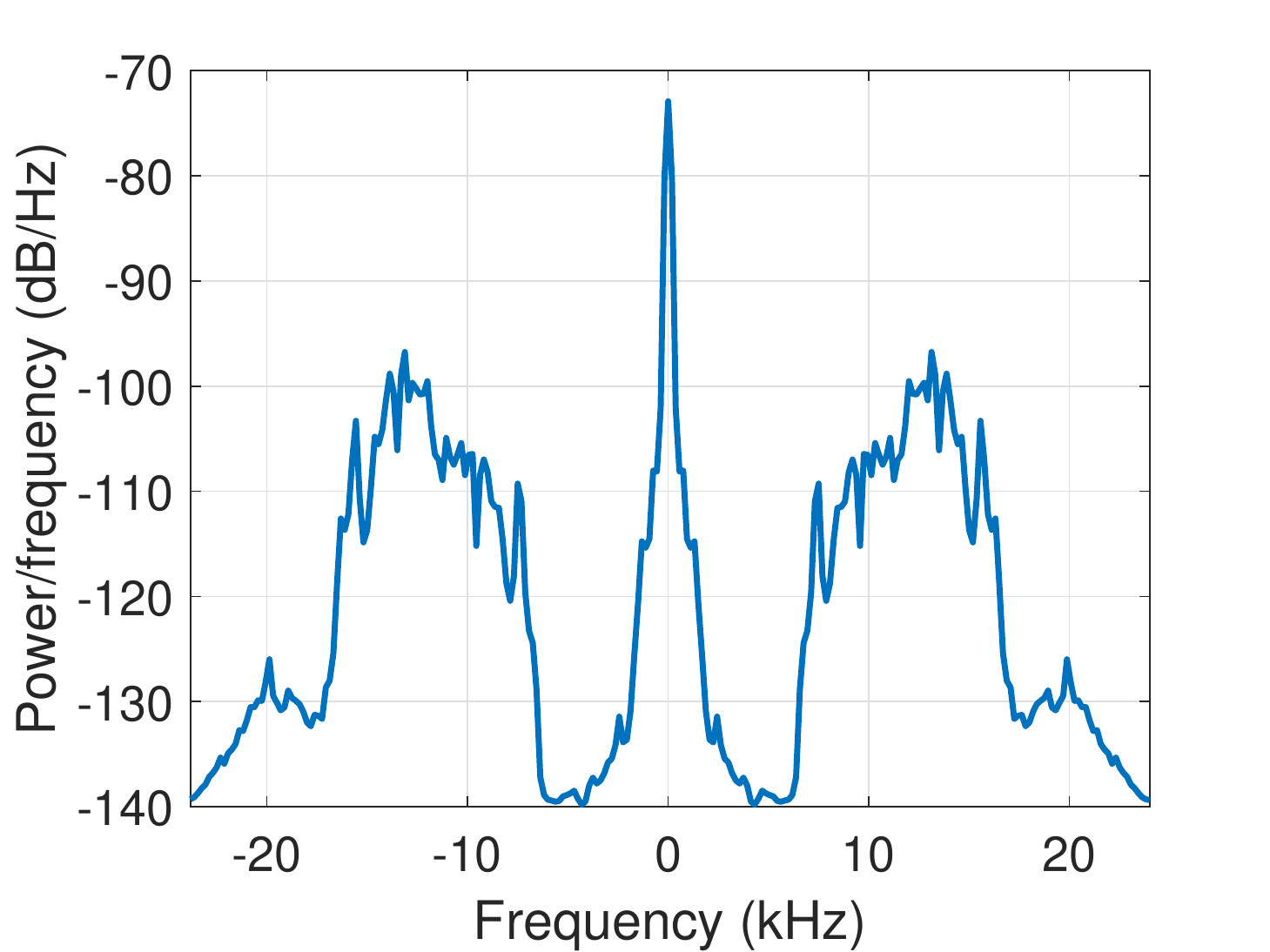} &
\hspace{-11mm}
\includegraphics[width=6.6cm
]{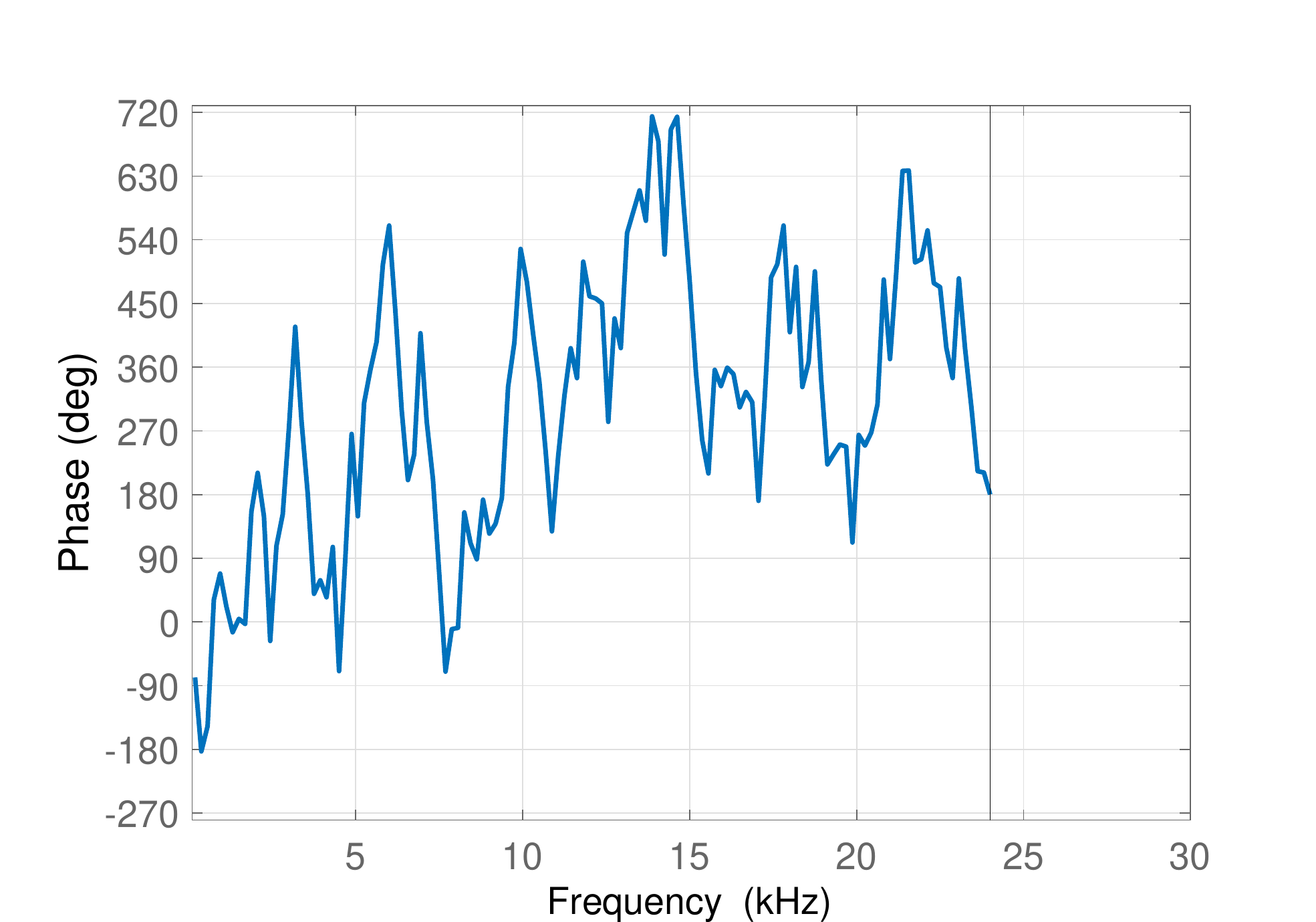} \\
\vspace{-1mm}
\small (c) &  \small (c)  & \small (c)\\
 \small (I) &  \small (II)  & \small (III)
\end{tabular}
\caption{\small Channel response while receiving the signal at (a)~Gateway at time 20:00; (b)~Gateway at time 21:00; (c)~Carol at time 20:00. Manta node deployed from the Ship was used as transmitter. The columns show (I)~Power versus frequency and time; (II)~Power spectrum density of the received signals; (III)~Phase variations versus frequency.}
\label{fig:channelG3fig}
\end{figure*}

\textbf{Results:} 
Figures~\ref{fig:channelG3fig} show the channel response at two sample receivers, i.e., Gateway at time 20:00 and 21:00 and Carol at time 20:00, while the Manta/Ship node was used as transmitter. In these figures, column~(I) shows the frequency spectrum of a sample received signals through different receivers for a specific duration. 
As shown in these figures, the received signals, which are recorded at different receivers (and also at different times), experience various channels. The effect of the amplitude response of the channel is reflected in the power spectral density of the received signals, as shown in Figs.~\ref{fig:channelG3fig}(II). Figures~\ref{fig:channelG3fig}(III) present the phase variations with respect to the frequency for these channels. 

\begin{figure*}[t]
\centering
\begin{tabular}{ccc}
\hspace{-0.45in}
\includegraphics[width=6.2cm
]{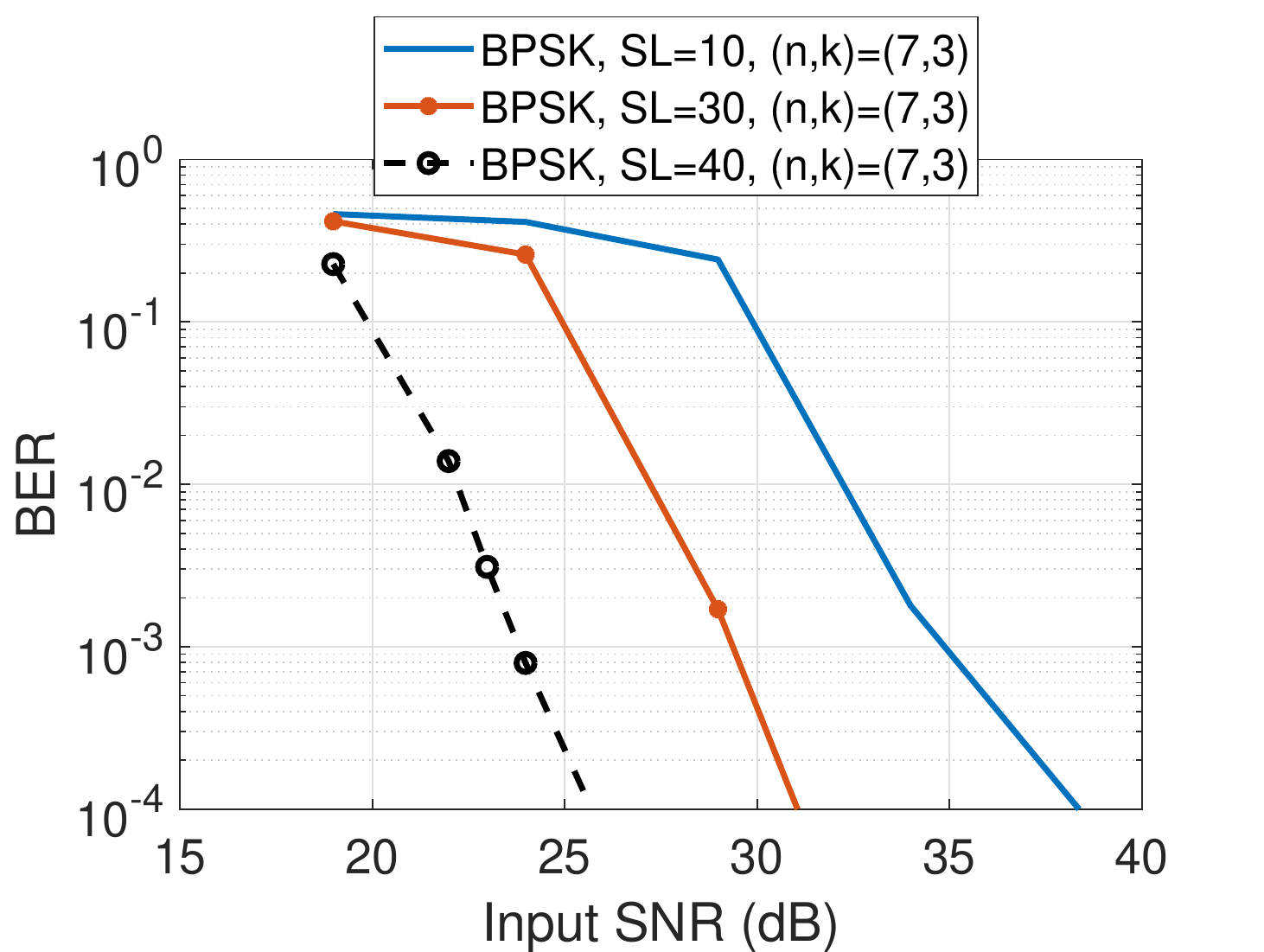} &  
\hspace{-8mm}
\includegraphics[width=6.2cm
]{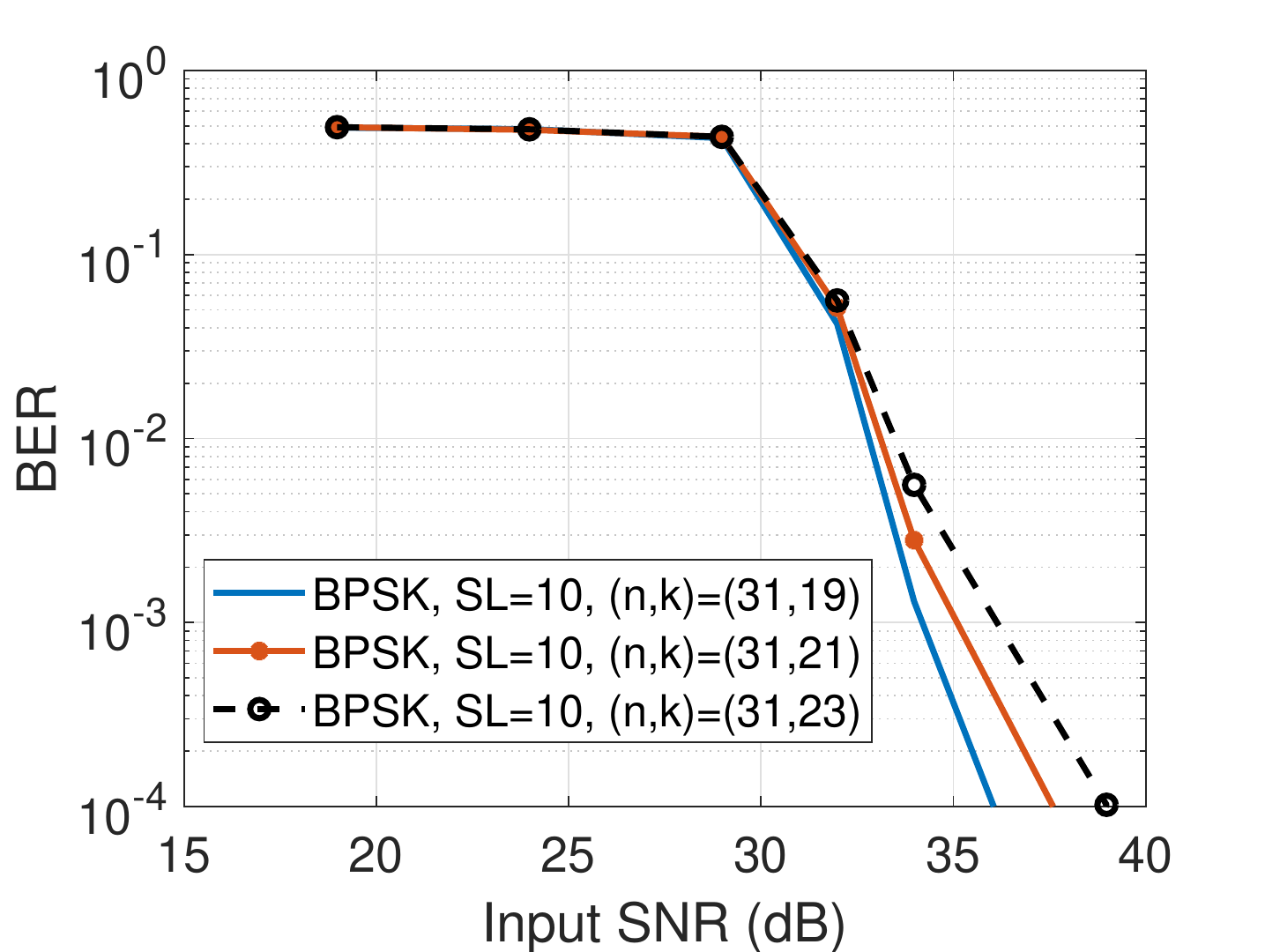} &  
\hspace{-8mm}
\includegraphics[width=6.2cm
]{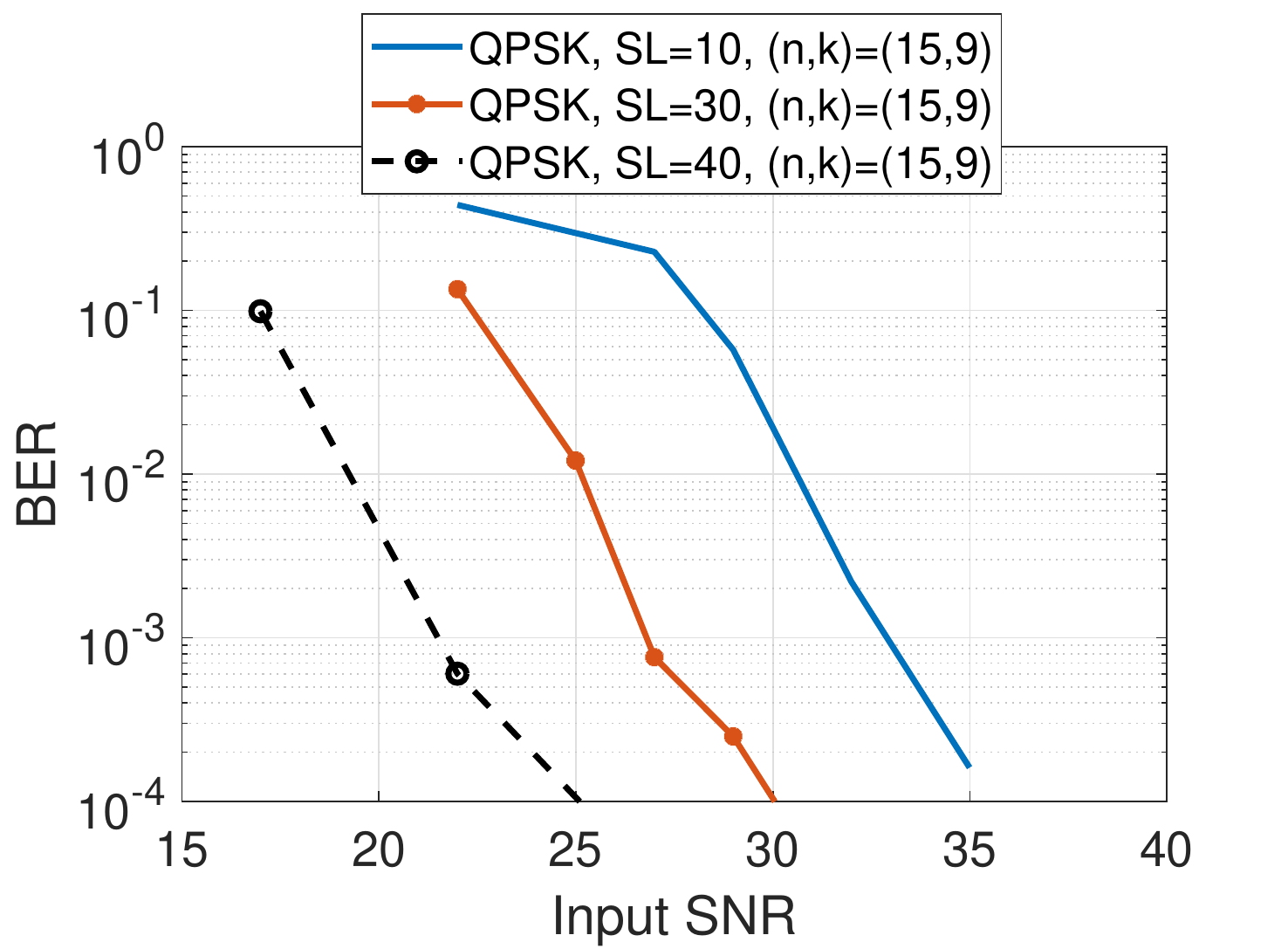} \\  
 \small (I) & \small (II)  & \small (III)
\end{tabular}
\caption{\small Single node transmission scenario; (I)~The Bit Error Rate~(BER) for different spreading lengths of the BPSK chaotic DS-CDMA signal; (II)~BER versus SNR shows the effect of coding strength in the BPSK chaotic DS-CDMA signal, (III)~BER for a RS $(15,9)$ coded QPSK with different spreading lengths.}
\label{fig:BER_1}
\end{figure*}
%
\begin{figure*}[!t]
\centering
\begin{tabular}{ccc}
\hspace{-0.45in}
\includegraphics[width=6.2cm
]{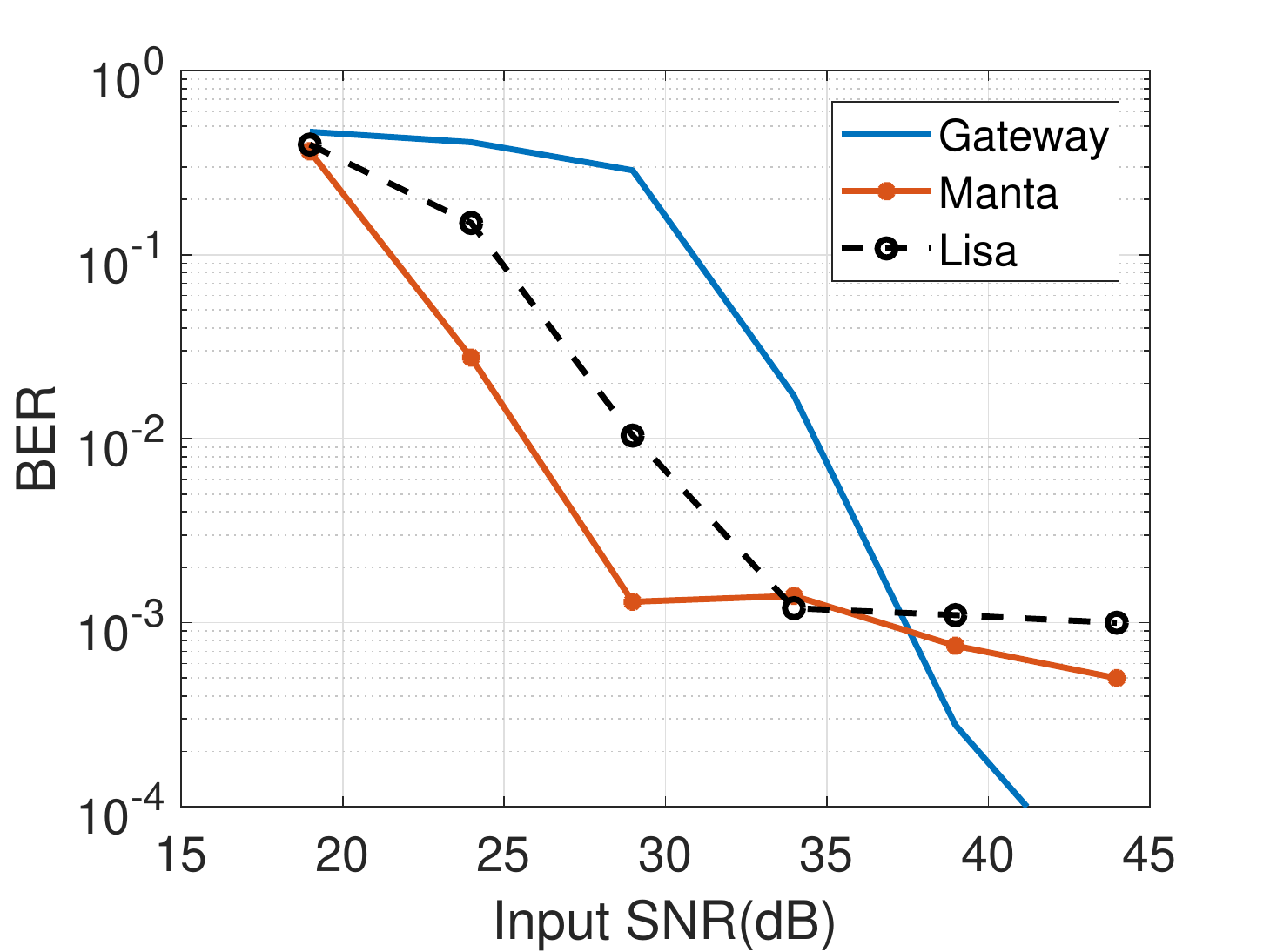} & 
\hspace{-8mm}
\includegraphics[width=6.2cm
]{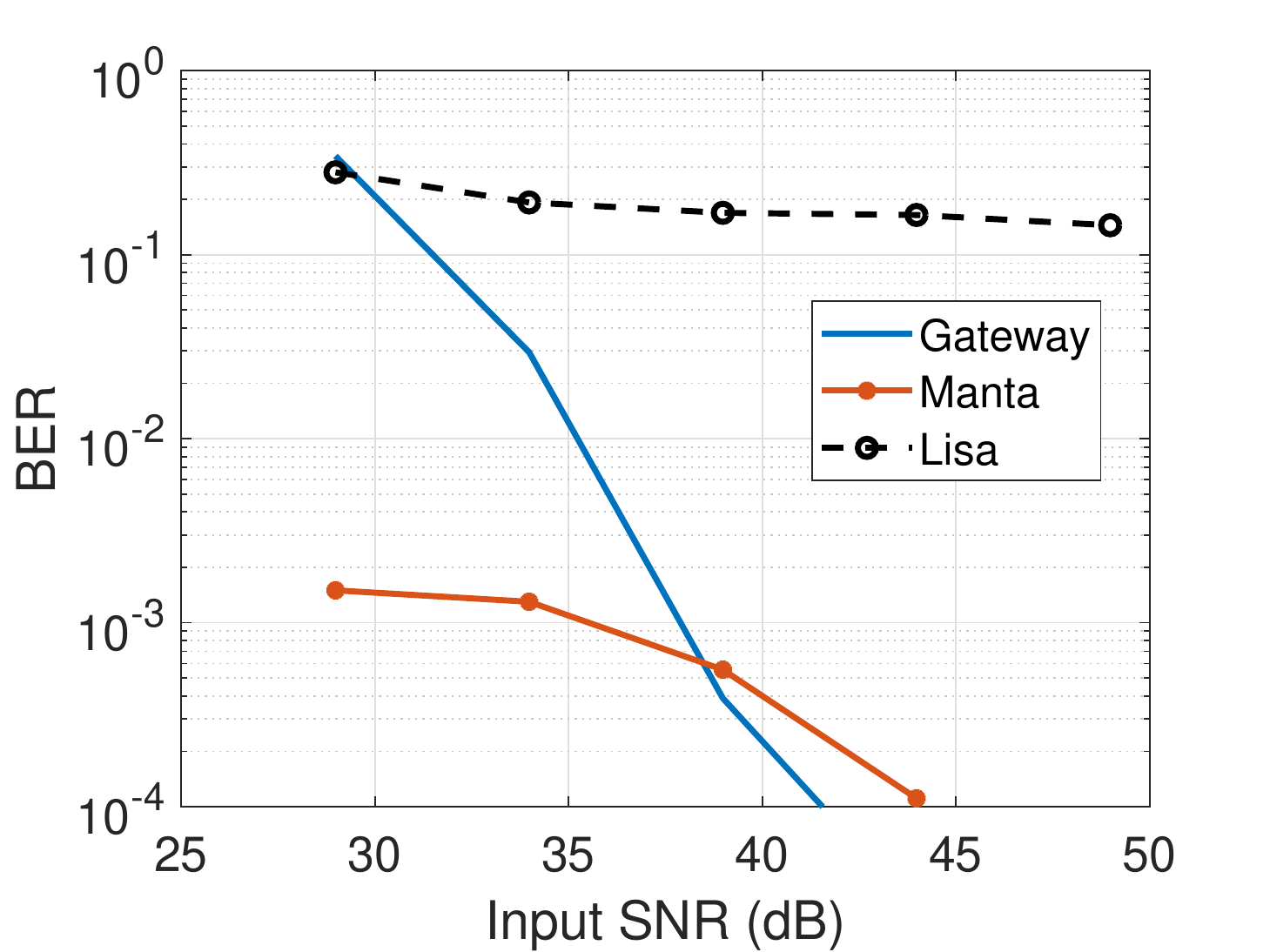} & 
\hspace{-8mm}
\includegraphics[width=6.2cm
]{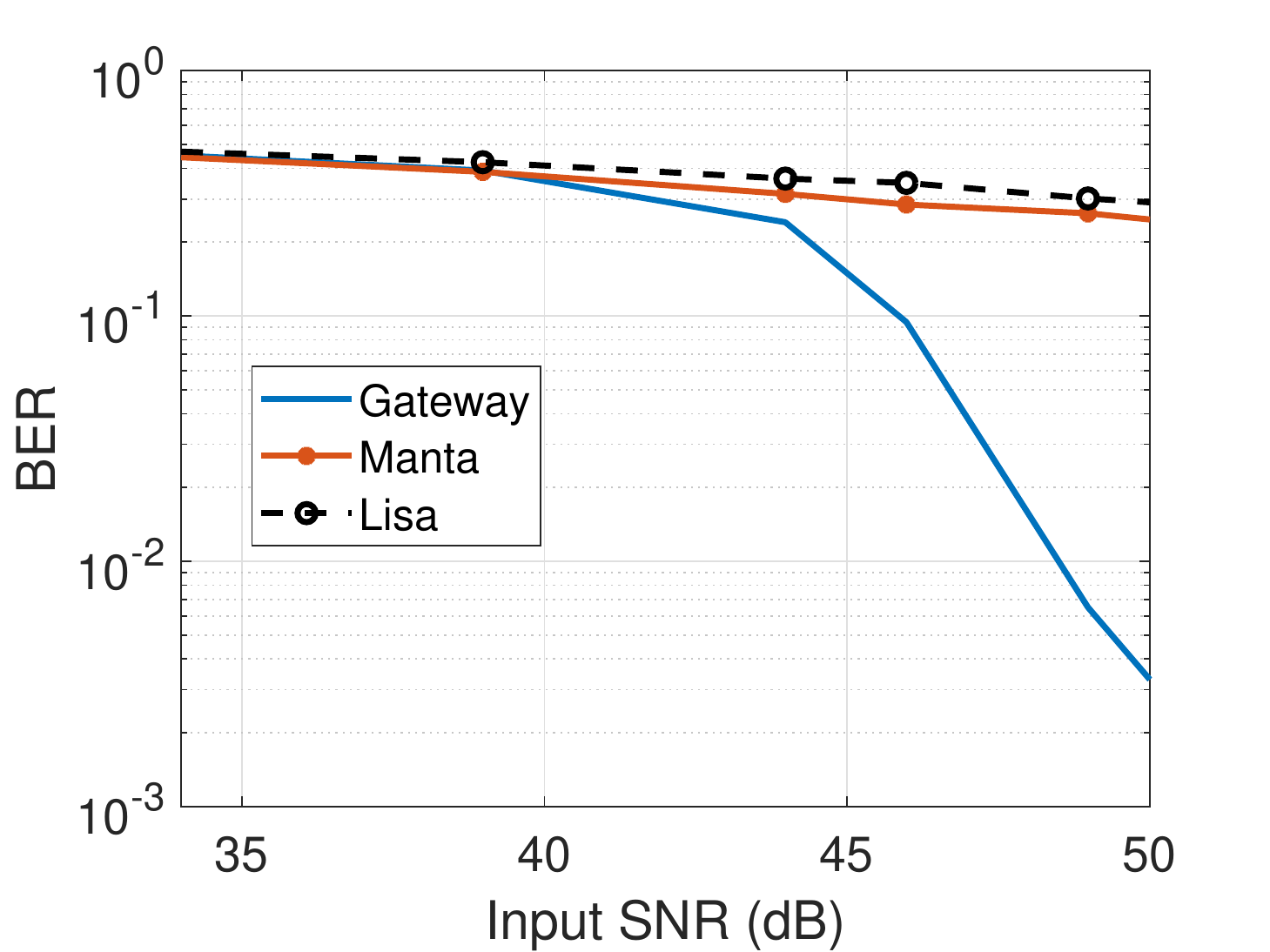} \\ 
 \small (I) & \small (II)  & \small (III)
\end{tabular}
\caption{\small BER in multiple-node transmission; (I)~Shows scenario $(a)$ in which all the three channels have a satisfactory quality; (II)~Shows scenario $(b)$ in which the signal which is transmitted from Lisa does not experience a good channel; (III)~Shows scenario $(c)$ in which the signals coming from both Lisa and Manta are impaired, but Gateway still has a channel with an acceptable quality.}
\label{fig:BER_2}
\end{figure*}

The experiments were performed for each setting with specific SNR. We scaled the data, applied power control, and added extra ambient noise at different noise levels to be able to present the performance in different SNRs for each experimented channel. Figures~\ref{fig:BER_1}(I)-(III) refer to the single node transmission scenario from the Gateway transmitter. In Fig.~\ref{fig:BER_1}(I), the BER for different spreading lengths for a BPSK signal is shown and confirmed that a higher spreading length leads to a lower BER. In Fig.~\ref{fig:BER_1}(II), the effect of coding rate on the performance was investigated. Changing the coding strength can improve the performance in high SNRs as shown in this figure. Figure~\ref{fig:BER_1}(III) shows the BER for a QPSK signal with a $(15,9)$ coding and different spreading lengths. 

\begin{figure*}
\centering
\begin{tabular}{ccc}
\hspace{-0.45in}
\includegraphics[width=6.2cm
]{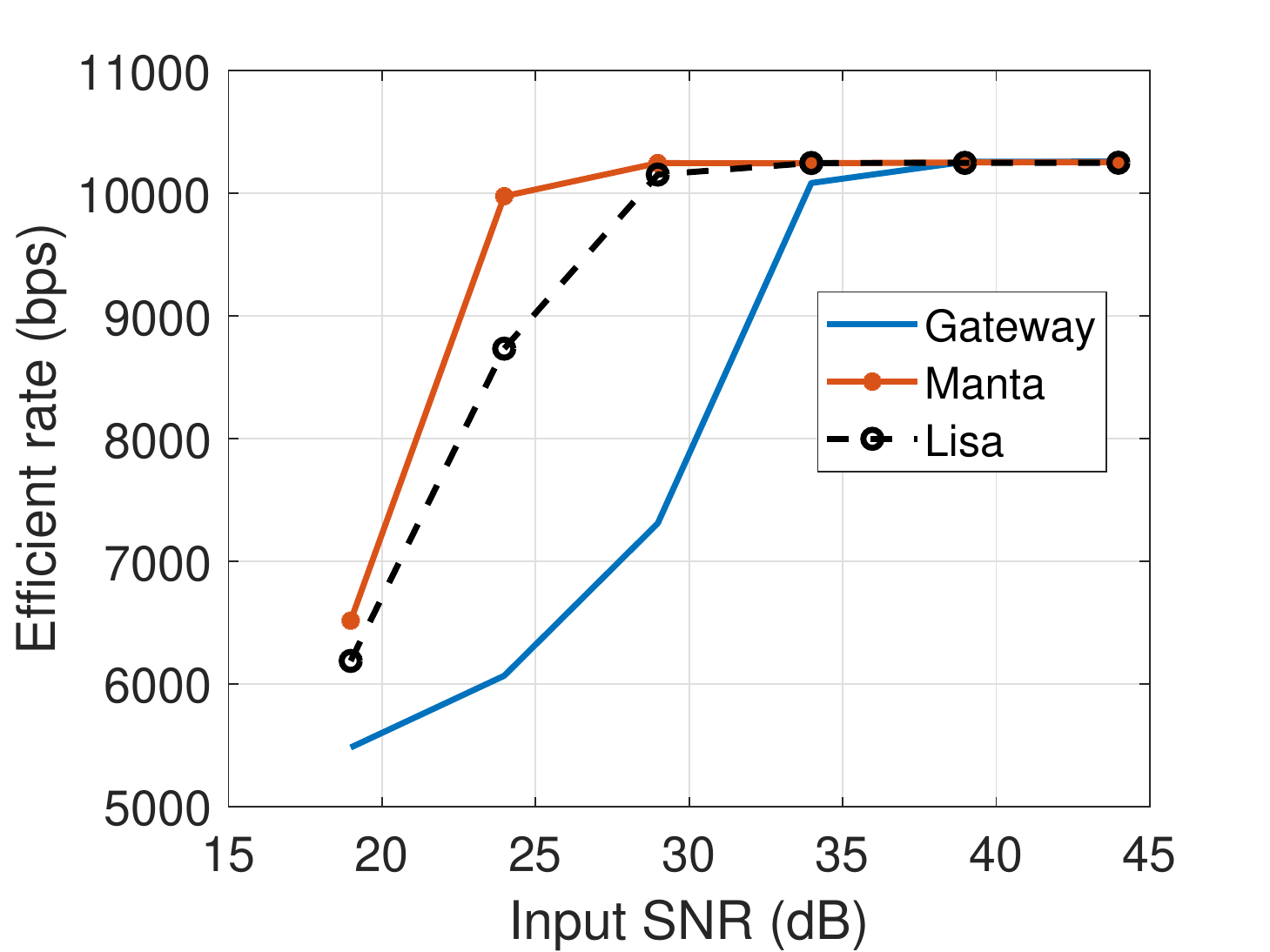} & 
\hspace{-8mm}
\includegraphics[width=6.2cm
]{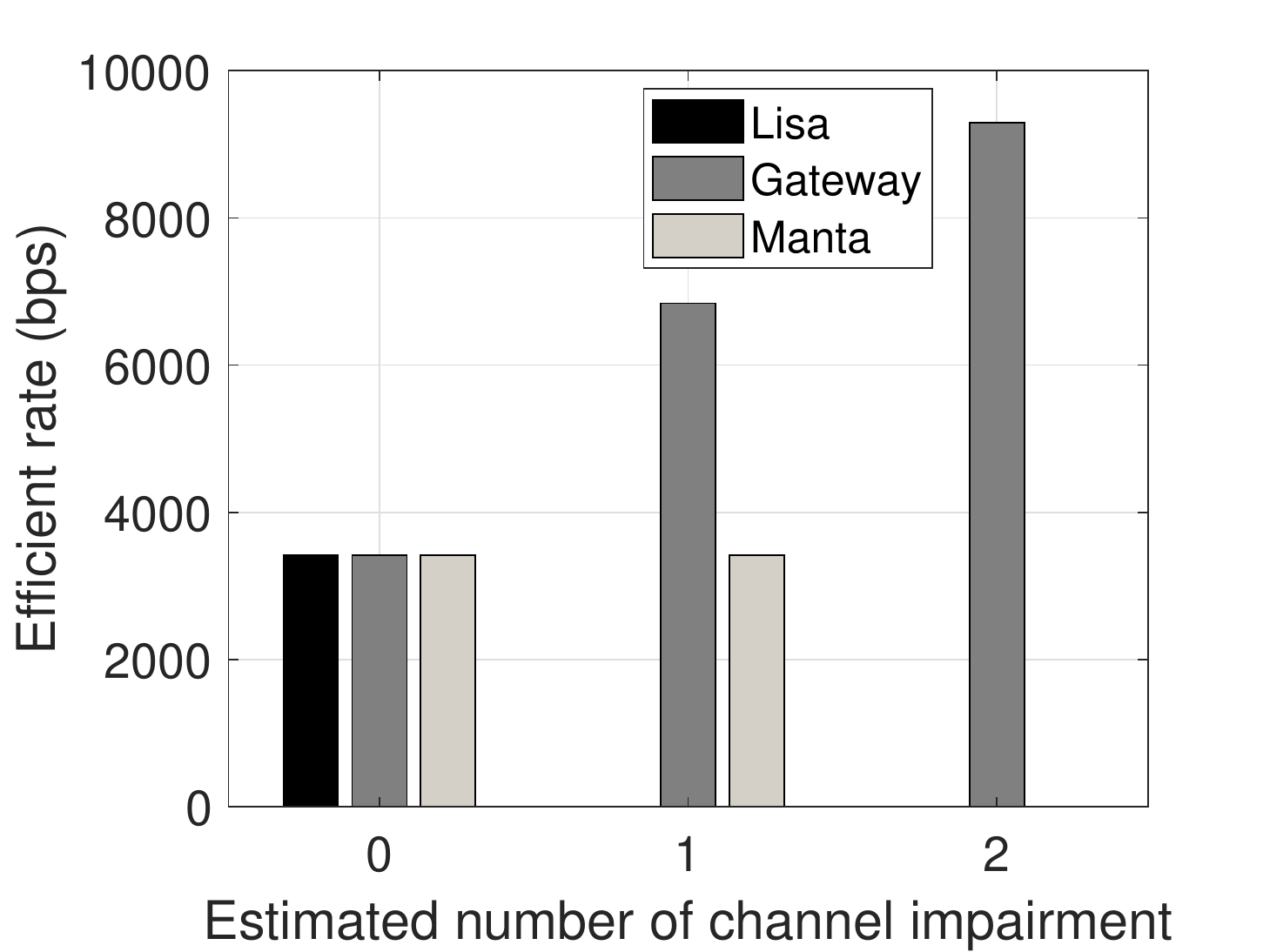} & 
\hspace{-8mm}
\includegraphics[width=6.2cm
]{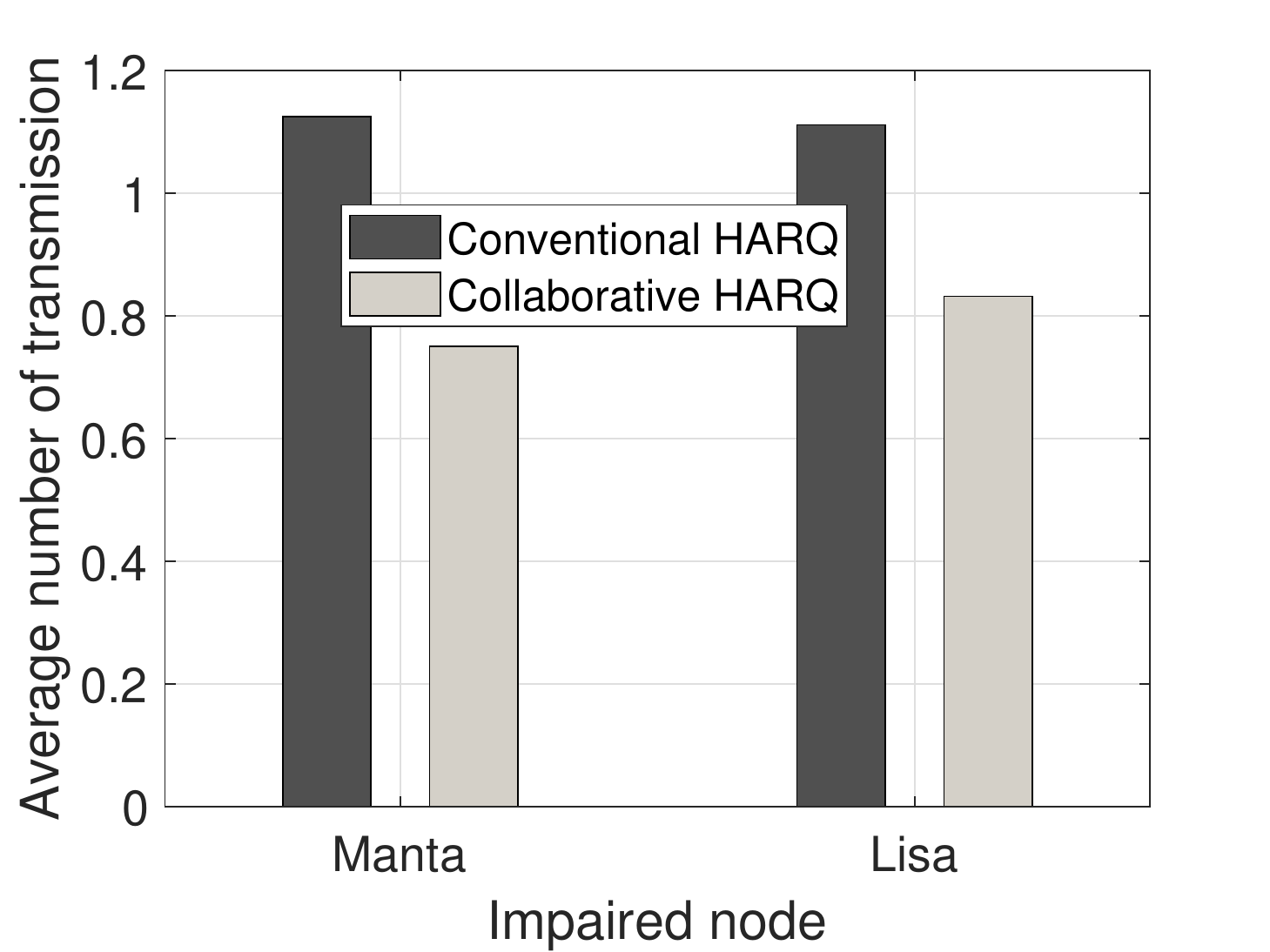} \\ 
 \small (I) & \small (II)  & \small (III)
\end{tabular}
\caption{\small (I)~Efficient rate when all three channels are reliable as described in scenario~(a); (II)~The efficient rate comparison and the collaboration in the different scenarios (a), (b), and (c); (III)~Average number of transmissions in the proposed collaborative solution versus the  conventional HARQ. Two cases have been compared; when Manta is impaired and Gateway collaborates and when Lisa is impaired and Manta is involved in the collaboration.}
\label{fig:RATE_1}
\end{figure*}
Figures~\ref{fig:BER_2}(I)-(III) show the BER in multiple-node transmission scenario, where all the nodes (i.e., Gateway, Manta, and Lisa) were transmitting simultaneously different signals. 
Figure~\ref{fig:BER_2}(I) shows scenario $(a)$ in which all three channels have a good quality and so there is no channel impairment, while Fig.~\ref{fig:BER_2}(II) represents a scenario $(b)$ in which the signal coming from Lisa does not experience a good channel; therefore, the network has one channel impairment. In this case, conventional HARQ will fail to deliver the data from this channel even with multiple retransmissions. The proposed collaborative solution will solve this problem, as discussed in Fig.~\ref{fig:RATE_1}. The other scenario, called $(c)$, is reported in Fig.~\ref{fig:BER_2}(III). This time two channel impairments are considered, both Lisa and Manta fail to deliver the data in the presence of a good channel from Gateway. Gateway will collaborate in our proposed solution to improve the total network efficiency, as reported in Fig.~\ref{fig:RATE_1}. Efficient rate is shown in Fig.~\ref{fig:RATE_1}(I) when all three channels are reliable as described in the scenario $(a)$. Figure~\ref{fig:RATE_1}(II) compares the efficient rate and the collaboration in the aforementioned scenarios. As an example, Gateway handles the Lisa's data in scenario $(b)$ and the whole network's data in scenario $(c)$. Figure~\ref{fig:RATE_1}(III) compares the average number of transmissions in the collaborative HARQ with the conventional one for two cases: $(i)$~the receiver returns a NACK to Manta. Gateway then collaborates with Manta in transmitting the extra redundancy which leads to a reduction in the average number of retransmissions as a result of this collaboration; $(ii)$~Lisa is and Manta is involved in the collaboration. By comparing the cases, we conclude that Gateway was a better collaborator in a comparable situation.

\begin{figure*}
\centering
\begin{tabular}{ccc}
\hspace{-0.45in}
\includegraphics[width=6.2cm
]{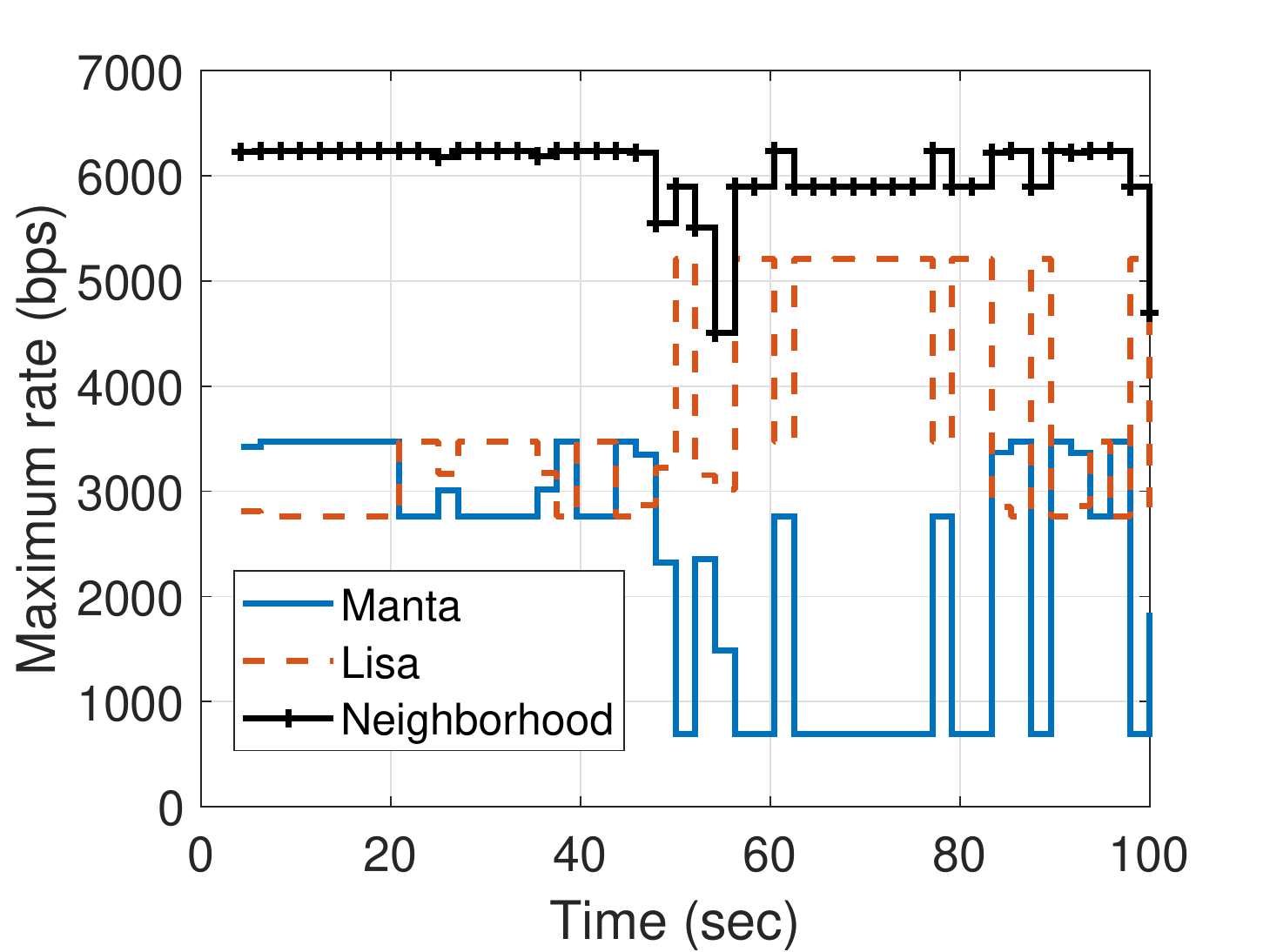} & 
\hspace{-8mm}
\includegraphics[width=6.2cm
]{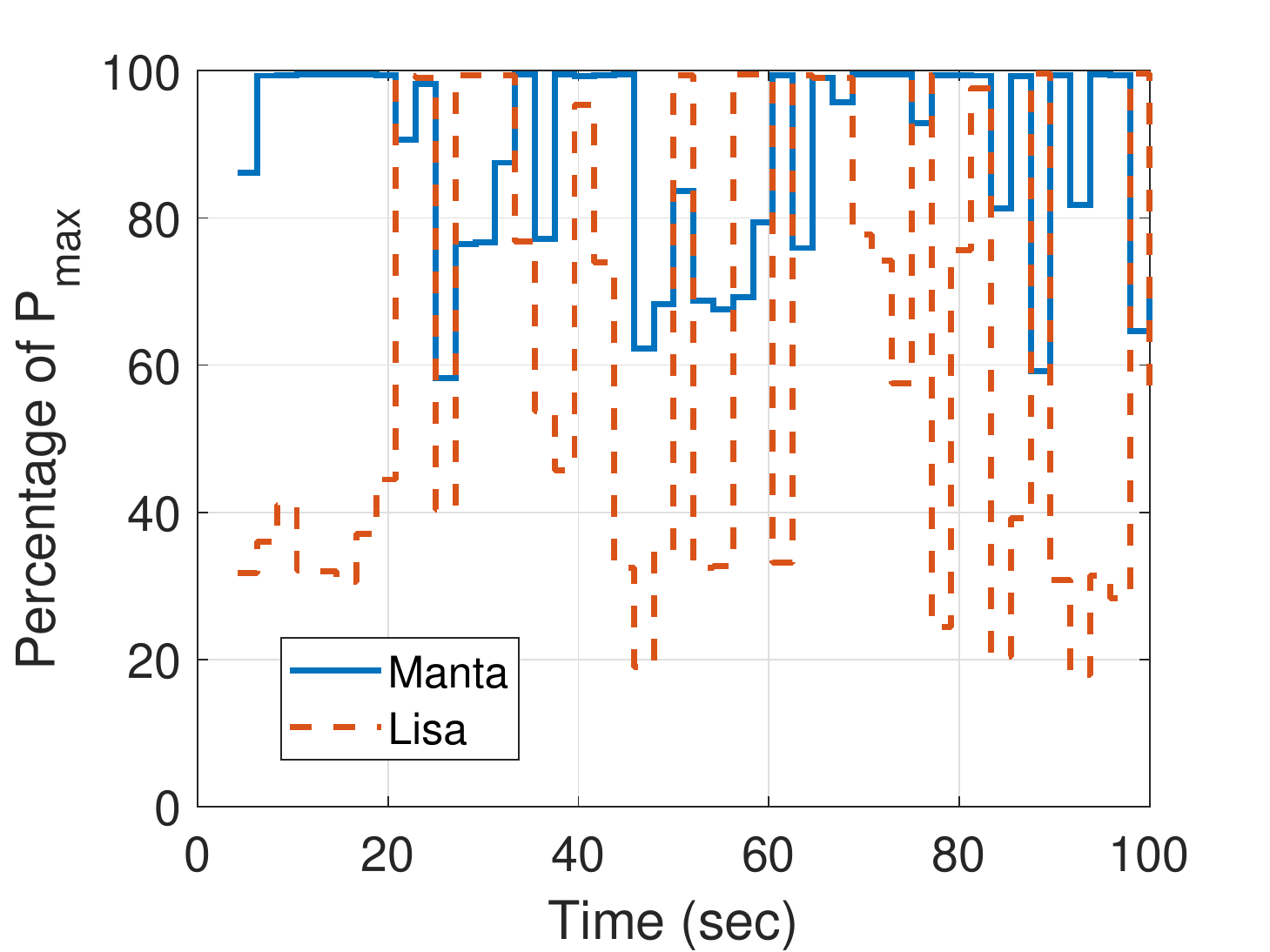} & 
\hspace{-8mm}
\includegraphics[width=6.2cm
]{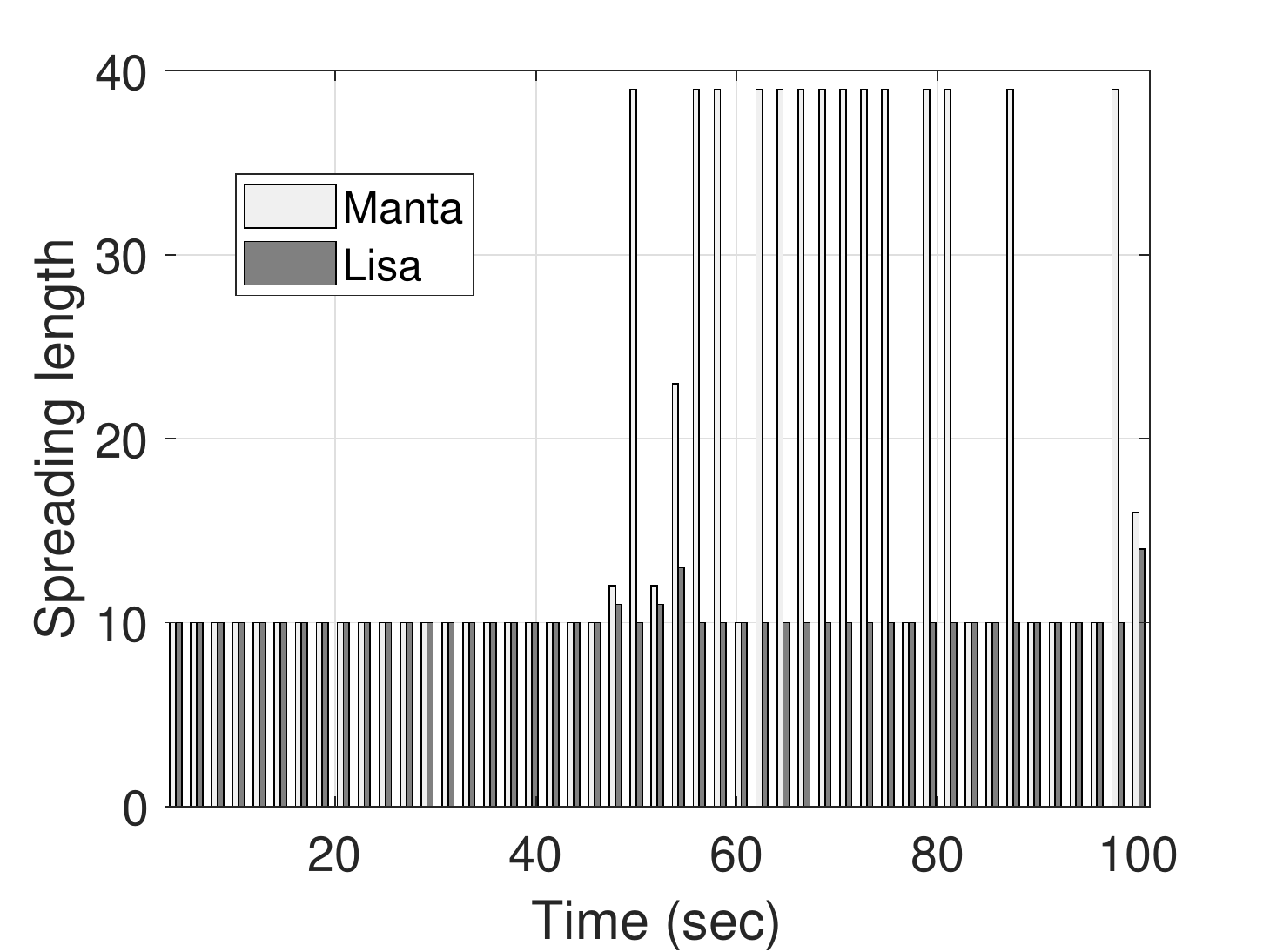} \\ 
 \small (I) & \small (II)  & \small (III)
\end{tabular}
\caption{\small When Manta and Lisa are two transmitters, this figure shows (I)~the maximum rate; (II)~transmit power percentage; (III)~the optimal CDMA spreading length, for a time period of $100\rm{s}$.}
\label{fig:RATE_2}
\end{figure*}
%
\begin{figure*}
\centering
\begin{tabular}{ccc}
\hspace{-0.45in}
\includegraphics[width=6.2cm
]{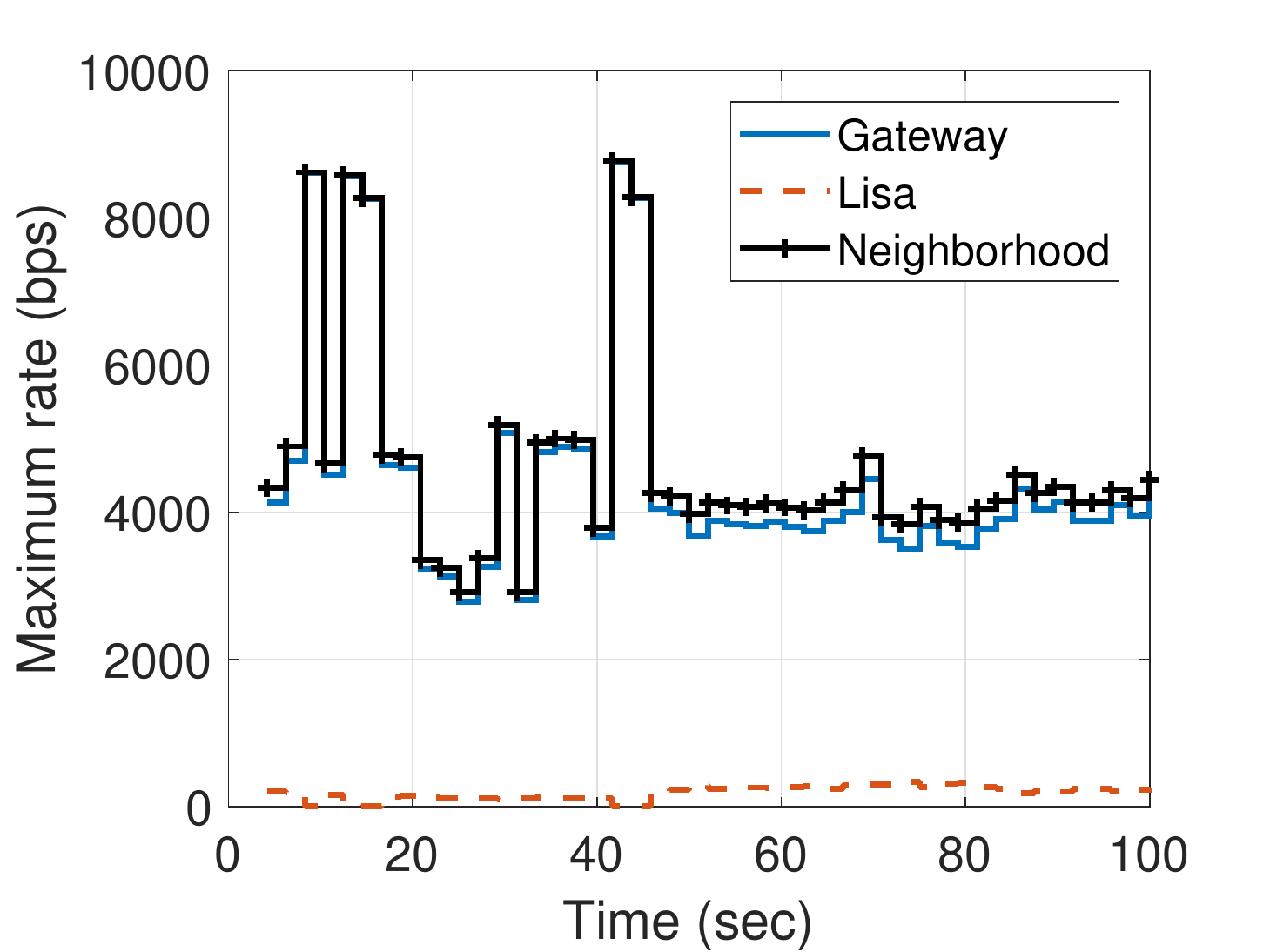} & 
\hspace{-8mm}
\includegraphics[width=6.2cm
]{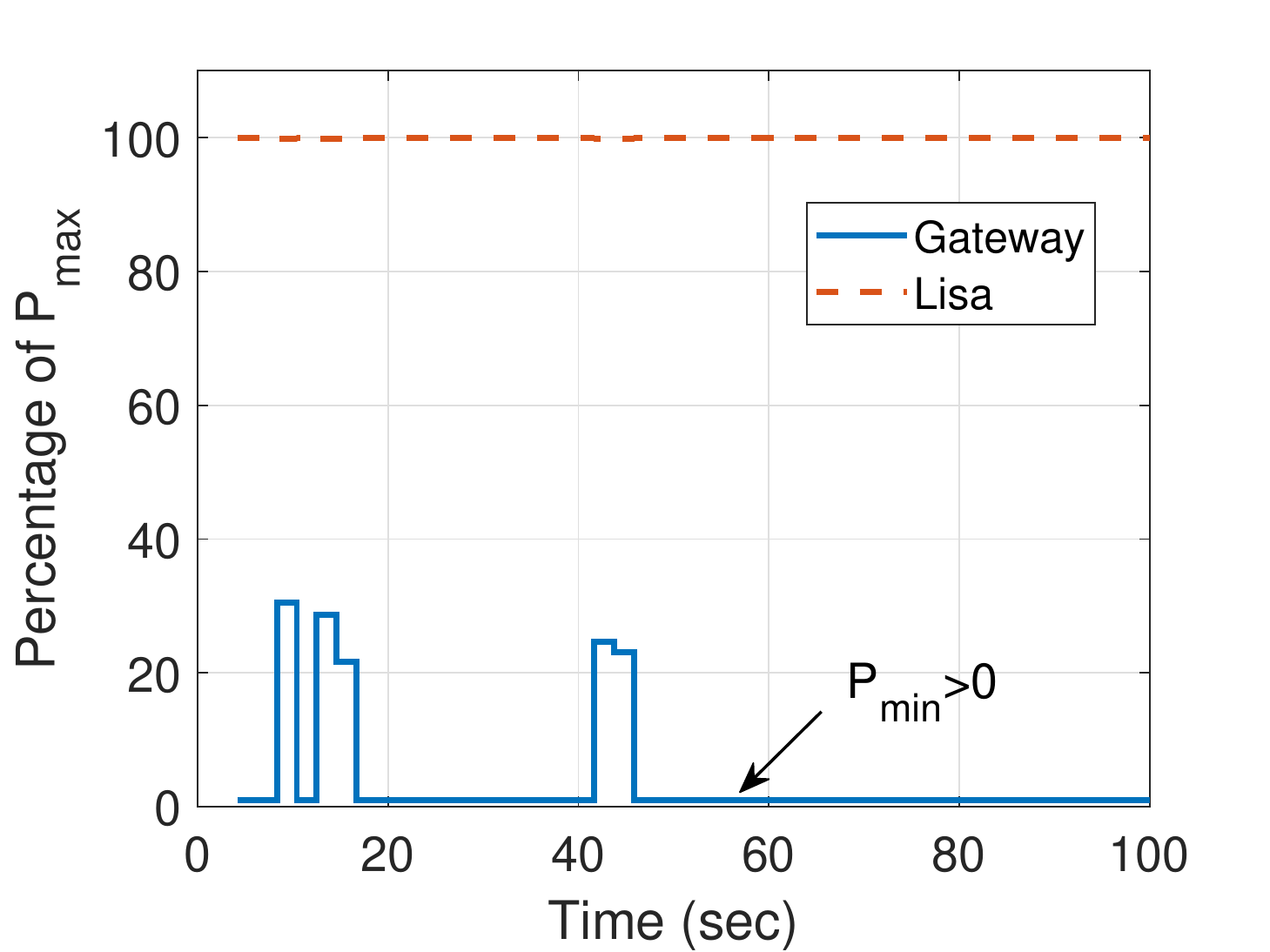} & 
\hspace{-8mm}
\includegraphics[width=6.2cm
]{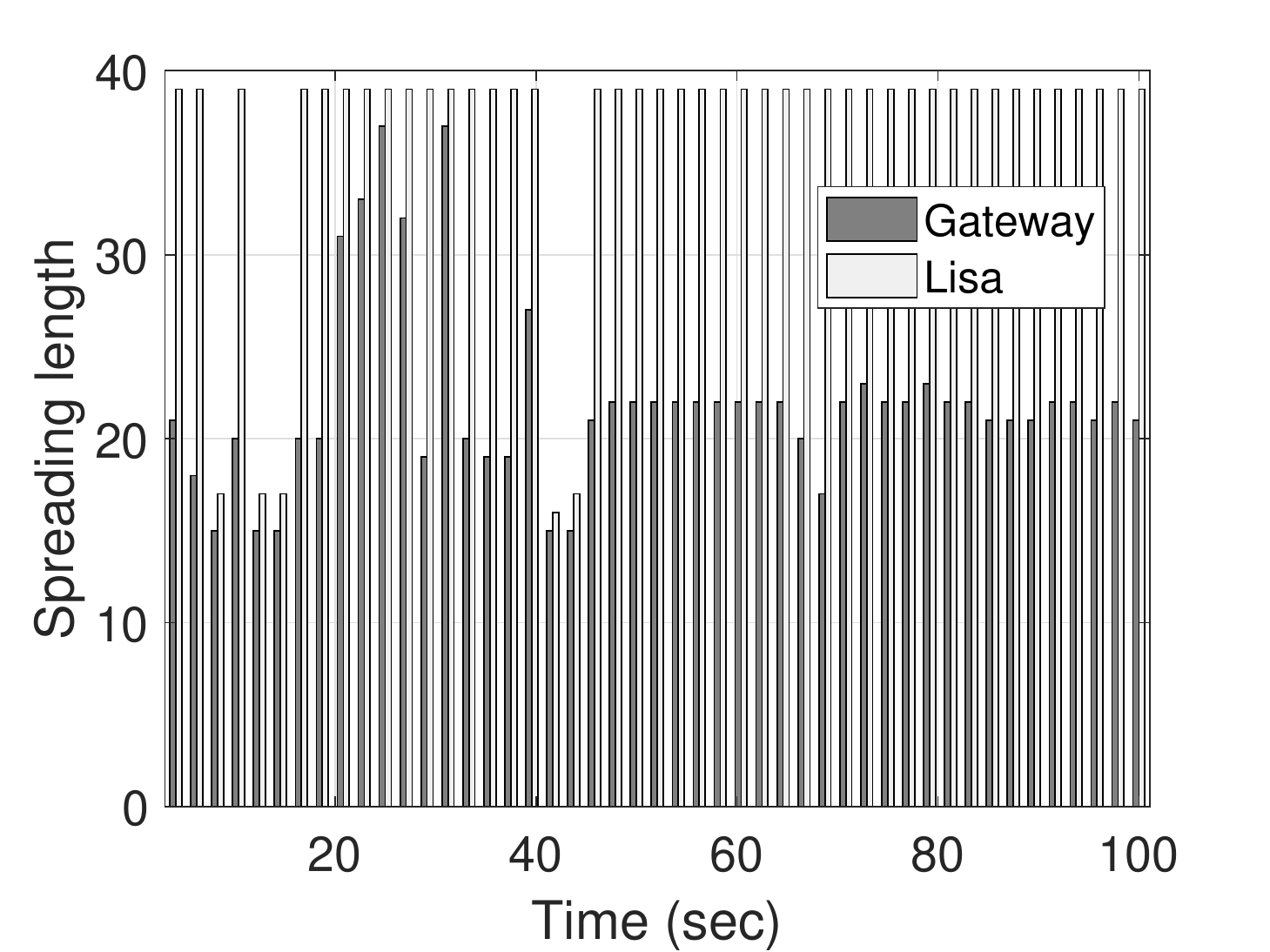} \\ 
 \small (I) & \small (II)  & \small (III)
\end{tabular}
\caption{\small When Gateway and Lisa are two transmitters, this figure shows (I)~the maximum rate; (II)~transmit power percentage; (III)~the optimal CDMA spreading length, for a time period of $100\rm{s}$.}
\label{fig:RATE_3}
\end{figure*}

To verify the energy efficiency of the proposed solution, we investigate the trade-offs between the maximum rate, power, and CDMA spreading length in Figs.~\ref{fig:RATE_2}-~\ref{fig:RATE_3}. We considered the experimental data from the channel of Lisa and Manta transmitters for a period of $100~\rm{s}$ and plotted the offline results of the optimization for maximum rate, transmit power, and the optimal CDMA spreading length, since the experiment was open loop. The goal is to maximize the neighborhood data rate which is shown in Fig.~\ref{fig:RATE_2}(I). In Fig.~\ref{fig:RATE_2}(II), the required power for each transmitter to reach this rate is plotted. This figure confirms the efficiency of power allocation since the transmitters transmit with a fraction of the maximum power. Figure~\ref{fig:RATE_2}(III) shows how the spreading length of these two transmitters adapts with the channel situation.

Figure~\ref{fig:RATE_3} shows the case in which Lisa and Gateway are the transmitters. Here, Gateway's signal is dominant and the signal that comes from Lisa experiences a poor channel. Lisa has to use its maximum power to defeat the interference coming from the strong signal of Gateway; however, it conveys a very low data rate. On the other hand, Gateway keeps its transmitter at the minimum power, while transmitting a great portion of data, as in Figs.~\ref{fig:RATE_3}(I)-(II). Our solution here is to switch off the node related to the weaker link, as described in the previous section. The other dominant node can handle the procedure with lower interference and better performance.

\begin{figure}[!t]
\centering 
\includegraphics[width=8cm
]{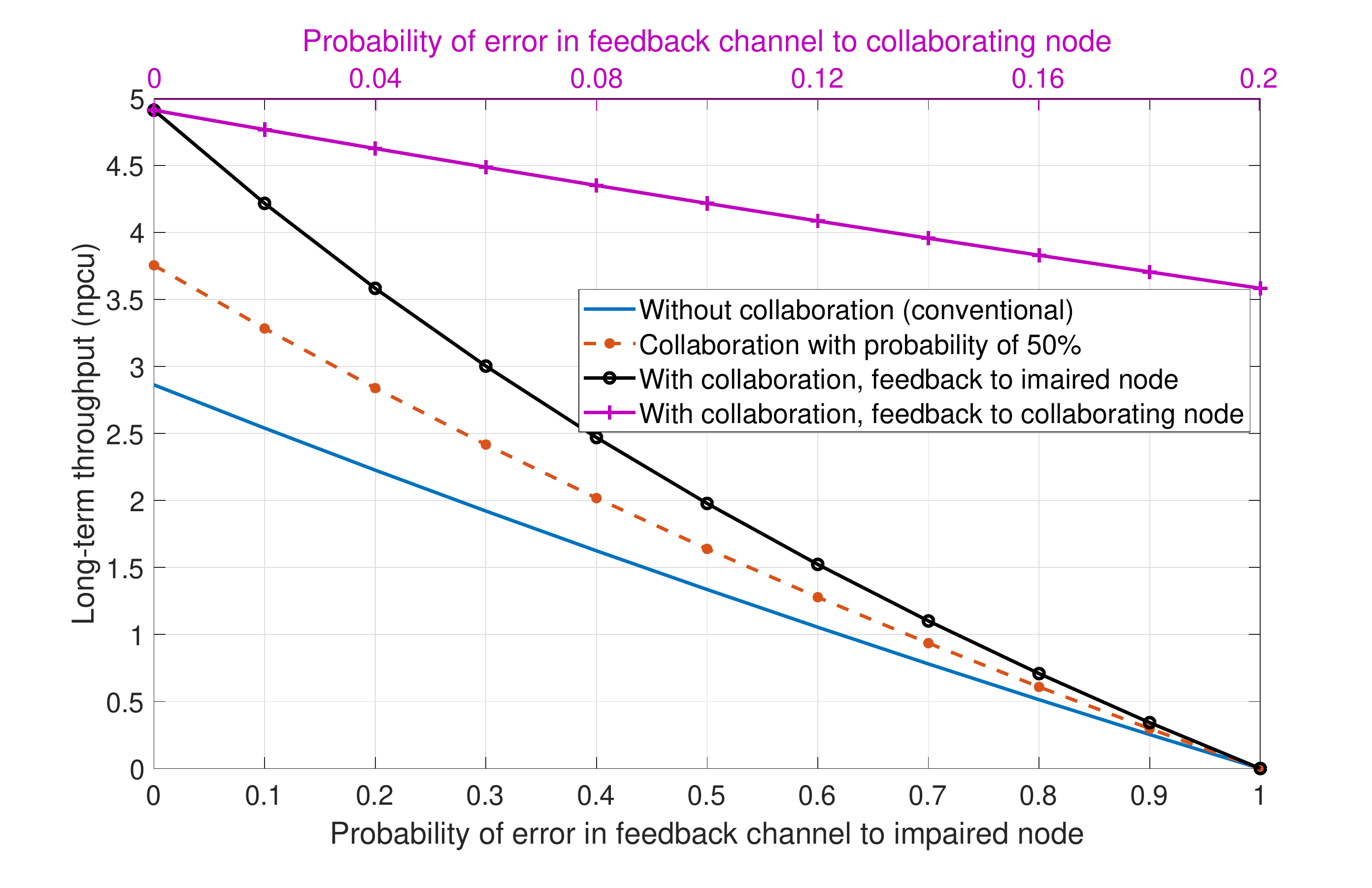} 
\caption{\small This long-term throughput simulation is illustrated for the feedback ACK/NACK messages under different assumptions. The case in which the impaired node does not collaborate, i.e., the conventional approach, is compared against the case in which collaboration occurs. The probability of collaboration with a neighbor is studied for $50\%$ and also for certain collaborations. When quality of the feedback channels to both the impaired and the collaborating nodes are similar is compared with the case when the collaborating node has a better quality.}\label{fig:feedbackerror}
\end{figure}

To evaluate the effect of errors in the feedback channel, a two-round HARQ simulation was investigated. Fig.~\ref{fig:feedbackerror} depicts the results in which two different assumptions were considered. Firstly, we assumed that the quality of feedback channel to both the impaired and collaborating nodes are similar in terms of probability of error. Secondly, we consider different probabilities of errors for those two feedback channels and simulated the long-term throughput for both cases. It is shown that the long-term throughput increases when the proposed algorithm is used.

\section{Conclusions}\label{sec:Con}
We introduced a collaborative strategy for a CDMA-based underwater Hybrid ARQ to increase the overall throughput of the network. Our solution leveraged both chaotic CDMA and HARQ properties to adjust the physical- and link-layer parameters and to compensate for the poor underwater acoustic communication links. System performance improvement and power control were considered, while the total throughput of the system was optimized. Our solution leverages both CDMA and HARQ properties to achieve multiple access and error protection in the scarce underwater bandwidth. Experimental data was first collected in a shallow-water configuration using the CMRE LOON testbed and processed to extend the results to other nodes via simulation. Additional data was then collected in a deeper-water scenario during the REP18-Atlantic sea-trial organized by CMRE, the Portuguese Navy~(PRT-N), and the Faculty of Engineering of the University of Porto (FEUP) to achieve a meaningful comparison under different conditions. As future work, we plan to implement our solution on a larger underwater network with heterogeneous nodes to manage a higher volume of data and to analyze the scalability of the solution. 

\textbf{Acknowledgments:} 
This work acknowledges the use of acoustic and environmental data acquired during the REP18-Atlantic sea trial, organized by the Portuguese Navy, NATO STO CMRE, and the Faculty of Engineering of the University of Porto. The authors also thank Konstantinos Pelekanakis and Giovanni Zappa at CMRE for their help during the experiments.

\bibliographystyle{IEEEtran}
\bibliography{ref_rev}

\normalsize

\end{document}